\documentstyle[a4p,12pt,epsfig]{article}

\begin{document}
\newcommand {\epem}     {e$^+$e$^-$}
\newcommand {\durham}   {$k_\perp$}
\newcommand {\rcone}    {$R$}
\newcommand {\econe}    {$\epsilon$}
\newcommand {\qq} {q$\overline{\mathrm{q}}$}
\newcommand {\gluglu} {gg}
\newcommand {\gincl} {g$_{\,\mathrm{incl.}}$}
\newcommand {\mngincl} {$\langle n_{\,\mathrm{ch.}} 
      \rangle_{\mathrm{g}_{\,\mathrm{incl.}}}$}
\newcommand {\mnuds} 
   {$\langle n_{\,\mathrm{ch.}} \rangle_{\mathrm{uds\,hemis.}}$}
\newcommand {\mnudsmod}
   {$\langle n_{\,\mathrm{ch.}} 
   \rangle_{\mathrm{uds\,hemis.}}^{41.8\,\mathrm{GeV}}$}
\newcommand {\egincl} {$\langle E 
      \rangle_{\mathrm{g}_{\,\mathrm{incl.}}}$}
\newcommand {\ecm} {\mbox{$E_{\mathrm{c.m.}}$}}
\newcommand {\lms} {$\Lambda_{\overline{\mathrm{MS}}}$}
\newcommand {\ejet} {$E_{\,\mathrm{jet}}$}
\newcommand {\rch} {r_{\,\mathrm{ch.}}}
\newcommand {\nch} {$n_{\,\mathrm{ch.}}$}
\newcommand {\mnch} {$\langle n_{\,\mathrm{ch.}} \rangle$}
\newcommand {\mnchmath} {\langle n_{\,\mathrm{ch.}} \rangle}
\newcommand {\lmsnffive} {$\Lambda_{\overline{\mathrm{MS}}}^{(n_f=5)}$}
\newcommand {\lmsnfthree} {$\Lambda_{\overline{\mathrm{MS}}}^{(n_f=3)}$}
\newcommand {\qzero} {$Q_0$}
\newcommand {\xee} {$x_{E}$}
\newcommand {\lnxee} {$\ln\,$({\xee})}
\newcommand {\ysph} {$y$}
\newcommand {\mysph} {$|y|$}
\newcommand {\mmysph} {|y|}
\newcommand {\rchysphone} {r_{\,\mathrm{ch.}}(\mmysph\leq 1)}
\newcommand {\rchysphtwo} {r_{\,\mathrm{ch.}}(\mmysph\leq 2)}
\newcommand {\pperp} {p_{\mathrm{T}}}
\newcommand {\rchsoft} {(r_{\,\mathrm{ch.}})^{p<4\,{\mathrm{GeV}}/c}
           _{0.8<\pperp<3\,{\mathrm{GeV}}/c}}
\newcommand {\rchsofttwogev} {(r_{\,\mathrm{ch.}})^{p<2\,{\mathrm{GeV}}/c}
           _{\pperp>0.8\,{\mathrm{GeV}}/c}}
\newcommand {\rchsoftpttwo} {(r_{\,\mathrm{ch.}})^{p<4\,{\mathrm{GeV}}/c}
           _{0.8<\pperp<2\,{\mathrm{GeV}}/c}}
\newcommand {\ptsoft} {p_{\mathrm{T}}\,(p\!<\!4~{\mathrm{GeV}}/c)  }
\newcommand {\cacf} {C$_{\mathrm{A}}$/C$_{\mathrm{F}}$}
\newcommand {\cca} {C$_{\mathrm{A}}$}
\newcommand {\ccf} {C$_{\mathrm{F}}$}
\newcommand {\nsig} {\mathrm{N}_{b/\sigma_{b}>2.5}}
\newcommand {\artwo} {\mbox{AR-2}}
\newcommand {\arthree} {\mbox{AR-3}}
\begin{titlepage}
\noindent
\begin{center}
  {\large   EUROPEAN LABORATORY FOR PARTICLE PHYSICS }
\end{center}

\bigskip\bigskip\bigskip
\begin{tabbing}
\` CERN-EP/99-028 \\
\` 24 February 1999 \\
\end{tabbing}

\bigskip\bigskip

\begin{center}{\Large\bf
Experimental properties of
gluon and quark jets \\
from a point source
}
\end{center}

\begin{center}
{\large
The OPAL Collaboration
}
\end{center}
\begin{center}{\large\bf  Abstract}\end{center}
\bigskip
\noindent
Gluon jets are identified in hadronic Z$^0$
decays as all the particles in a hemisphere
opposite to a hemisphere containing
two tagged quark jets.
Gluon jets defined in this manner
are equivalent to gluon jets
produced from a color singlet point source and
thus correspond to the definition employed
for most theoretical calculations.
In a separate stage of the analysis,
we select quark jets in a manner
to correspond to calculations,
as the particles in hemispheres of
flavor tagged light quark (uds) events.
We present the distributions of rapidity, scaled energy,
the logarithm of the momentum,
and transverse momentum with respect to the jet axes,
for charged particles in these gluon and quark jets.
We also examine the charged particle multiplicity distributions
of the jets in restricted intervals of rapidity.
For soft particles at large $\pperp$,
we observe the charged particle
multiplicity ratio of gluon to quark jets to be
$2.29\pm0.09\,{\mathrm{(stat.)}}
\pm0.15\,{\mathrm{(syst.)}}$,
in agreement with the prediction that
this ratio should approximately equal the ratio of
QCD color factors, {\cacf}$\,$=$\,$2.25.
The intervals used to define soft particles 
and large $\pperp$ for this result,
$p$$\,<\,$4~GeV/$c$
and 0.8$\,<\,$$\pperp$$\,<\,$3.0~GeV/$c$,
are motivated by the predictions of the Herwig Monte Carlo
multihadronic event generator.
Additionally,
our gluon jet data allow a sensitive
test of the phenomenon of non-leading QCD
terms known as color reconnection.
We test the model of color reconnection implemented in
the Ariadne Monte Carlo multihadronic event generator
and find it to be disfavored by our data.

\bigskip\bigskip\bigskip
\begin{center}{\large
(To be submitted to Eur. Phys. Jour. C)
}\end{center}

\end{titlepage}

{\small
\begin{center}{
G.\thinspace Abbiendi$^{  2}$,
K.\thinspace Ackerstaff$^{  8}$,
G.\thinspace Alexander$^{ 23}$,
J.\thinspace Allison$^{ 16}$,
N.\thinspace Altekamp$^{  5}$,
K.J.\thinspace Anderson$^{  9}$,
S.\thinspace Anderson$^{ 12}$,
S.\thinspace Arcelli$^{ 17}$,
S.\thinspace Asai$^{ 24}$,
S.F.\thinspace Ashby$^{  1}$
% ...
}\end{center}\bigskip
\bigskip
$^{  1}$School of Physics and Astronomy, University of Birmingham,
Birmingham B15 2TT, UK
\newline
$^{  2}$Dipartimento di Fisica dell' Universit\`a di Bologna and INFN,
I-40126 Bologna, Italy
\newline
$^{  3}$Physikalisches Institut, Universit\"at Bonn,
D-53115 Bonn, Germany
\newline
% ...
$^{  5}$Cavendish Laboratory, Cambridge CB3 0HE, UK
\newline
% ...
$^{  8}$CERN, European Organisation for Particle Physics,
CH-1211 Geneva 23, Switzerland
\newline
$^{  9}$Enrico Fermi Institute and Department of Physics,
University of Chicago, Chicago IL 60637, USA
\newline
% ...
$^{ 12}$Indiana University, Department of Physics,
Swain Hall West 117, Bloomington IN 47405, USA
\newline
% ...
$^{ 16}$Department of Physics, Schuster Laboratory, The University,
Manchester M13 9PL, UK
\newline
$^{ 17}$Department of Physics, University of Maryland,
College Park, MD 20742, USA
\newline
% ...
$^{ 23}$Department of Physics and Astronomy, Tel Aviv University,
Tel Aviv 69978, Israel
\newline
$^{ 24}$International Centre for Elementary Particle Physics and
Department of Physics, University of Tokyo, Tokyo 113-0033, and
Kobe University, Kobe 657-8501, Japan
\newline
% ...
$^{  c}$ and Institute of Nuclear Research, Debrecen, Hungary
\newline
% ...
\newline
}

\clearpage\newpage
\section{Introduction}
\label{sec-intro}

Gluon jets have been a subject of intensive experimental
investigation since the time
of their first observation~\cite{bib-petra}.
It has proven difficult to obtain theoretically
meaningful information about the internal
properties of gluon jets, however,
due to the experimental difficulty
of identifying gluon jets in a
manner which corresponds to calculations in
Quantum Chromodynamics (QCD).
The theoretical description of gluon jets usually relies
on the creation of a gluon jet pair, {\gluglu},
from a color singlet point source,
allowing an unambiguous definition of the gluon jet's properties
by summing inclusively over the particles in
an event hemisphere.
Point-source creation of a high energy {\gluglu} pair
({\ejet}$\,>\,$~5~GeV)
is not a process which has been observed in 
nature, however.\footnote{It is possible
to identify a pure source of {\gluglu} events in
radiative $\Upsilon$ decays~\cite{bib-cleoqg};
however,
the jet energies are only about 5~GeV in this case,
which limits their usefulness for jet studies.}
Instead, experimenters have relied on jet finding algorithms
to isolate high energy gluon jets within other types of events.
At {\epem} colliders,
most studies of gluon jets employ
a jet finding algorithm to
select a sample of three-jet {\qq}g events.
The same jet finder is used
to divide the particles
of an event into a gluon jet part and two quark jet parts.
At hadron colliders,
jet finding algorithms are used to
select two-jet {\gluglu} events which
do {\it not} arise from a point source
since the gluon jets
are color-connected to other jets and to the
underlying event from the proton remnants.
Jet finders are used to artificially divide events into
gluon and non-gluon jet parts similarly to
the {\epem} case.
The results obtained for the gluon jet properties
at either type of collider
are found to depend strongly 
on the jet finding algorithm and,
as a consequence,
have limited theoretical significance.

In~\cite{bib-jwg},
a method was introduced to experimentally
identify gluon jets in a manner
which yields a close correspondence to the
theoretical definition.
The method is based on rare events
of the type {\epem}$\rightarrow\,${\qq}$\,${\gincl},
in which the q and $\overline{\mathrm{q}}$ are
identified quark jets\footnote{In this analysis
we make no distinction between quark and antiquark jets
and refer to both as ``quark'' jets.}
which appear in the same hemisphere of an 
{\epem} multihadronic annihilation event.
The object {\gincl}, taken to be the gluon jet,
is defined by all particles observed in the
hemisphere opposite to that containing 
the q and~$\overline{\mathrm{q}}$.
The properties of gluon jets found using this method
are almost entirely independent of the choice of the
jet finding algorithm used to define the quark jets.
In the limit that the q and~$\overline{\mathrm{q}}$ 
are collinear,
the gluon jets {\gincl} are produced under the same
conditions as gluon jets in 
{\gluglu} events~\cite{bib-khoze}.
The {\gincl} jets therefore correspond closely to 
single gluon jets in {\gluglu} events,
defined by dividing the {\gluglu} events in half
using the plane perpendicular to the principal event~axis.

In several previous studies~\cite{bib-opalqg96,bib-opalqg97},
we employed the {\gincl} hemisphere method
of defining gluon jets to determine the charged particle
multiplicity distribution of the jets.
The data were collected using the OPAL detector at 
the {\epem} collider LEP at CERN.
In this paper,
we extend our investigation of gluon jets
to other distributions,
in particular to rapidity, the logarithm of momentum,
transverse momentum with respect to the jet axis,
scaled energy, and multiplicity
in restricted rapidity intervals,
for charged particles in the jets.
The results for gluon jets are compared to those
of light flavored (uds) quark jets.
We define a uds jet to be
all the particles in a hemisphere of an
{\epem}$\rightarrow\,$Z$^0\rightarrow\,$$hadrons$
event in which the Z$^0$ decays into a quark pair
{\qq} with q$\,$=$\,$u, d or~s.
Use of light quark events results in
a better correspondence between the data
and the massless quark assumption
employed for most theoretical calculations.
Use of event hemispheres to define the quark jets
yields an inclusive definition analogous to that
of the gluon jets.

\begin{figure}[ht]
\epsfxsize=17cm
\epsffile[25   150 555 750]{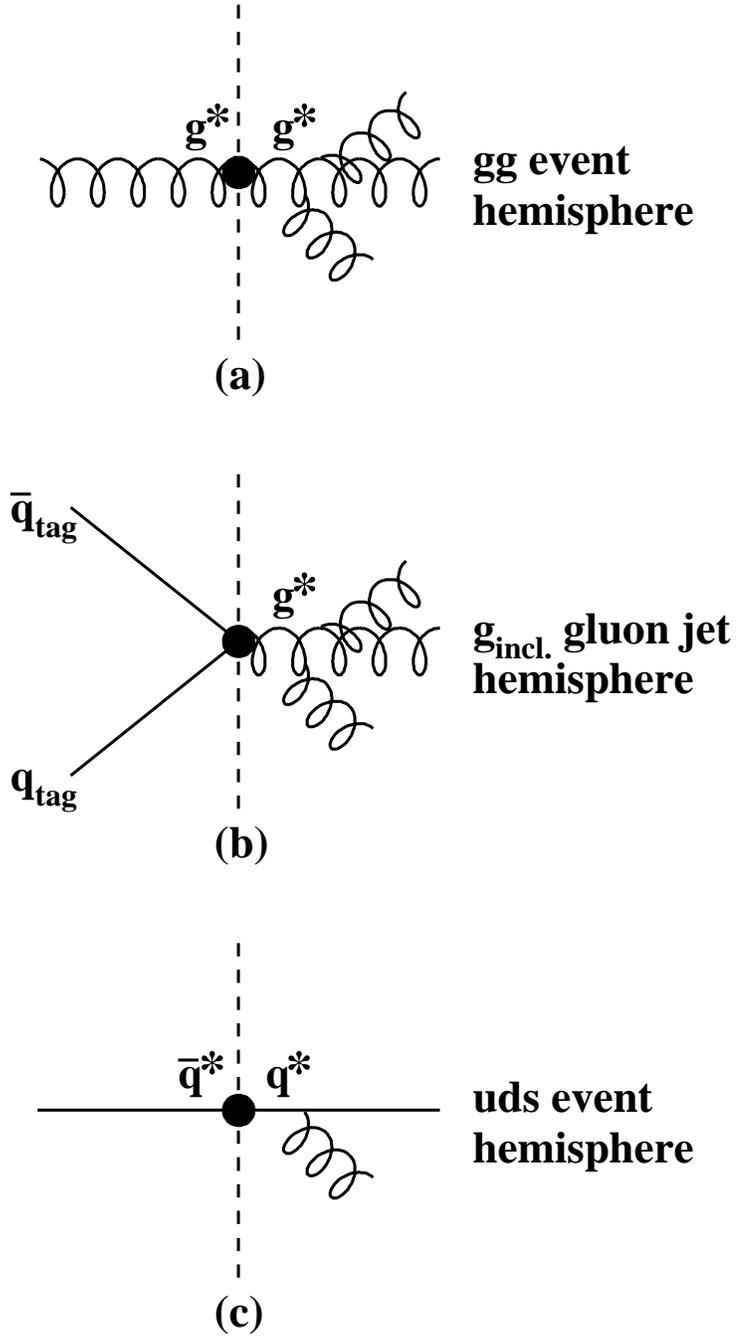}
\caption{Event types pertinent
to this analysis.
The dashed vertical lines represent
hemisphere boundaries,
defined in our study by the plane perpendicular
to the thrust axis,
while the large solid dots represent
a color singlet point source.
(a)~{\gluglu} production.
(b)~{\epem}$\rightarrow\,$q$_{\mathrm{tag}}
\overline{\mathrm{q}}_{\mathrm{tag}}${\gincl}:
The quark jets q$_{\mathrm{tag}}$ and
$\overline{\mathrm{q}}_{\mathrm{tag}}$ are
tagged b jets defined using a jet algorithm
and are used only as a tool to identify
the {\gincl} jet hemispheres.
The {\gincl} jet hemispheres provide the
gluon jet sample for our study.
The {\gincl} jets yield virtually the
same results for the experimental observables in
our study as the hemispheres of {\gluglu} events
shown in part~(a).
(c)~{\epem}$\rightarrow\,$q$\overline{\mathrm{q}}$,
with q a light (uds) quark:
Hemispheres in these events provide the
quark jet sample for our study.
}
   \label{fig-diagrams}
\end{figure}

Fig.~\ref{fig-diagrams} illustrates the three 
types of event pertinent to our study.
Fig.~\ref{fig-diagrams}a shows a diagram for 
{\gluglu} production from a color singlet point source.
The production of {\gincl} jets in {\epem}
annihilations,
providing an experimentally accessible source
of high energy gluon jets with nearly identical 
properties to the gluon jets in {\gluglu} events,
is shown in Fig.~\ref{fig-diagrams}b.
Last,
Fig.~\ref{fig-diagrams}c
shows uds jet production in {\epem} annihilations.

A topic of recent interest is that of 
color reconnection~\cite{bib-rcon}.
The phenomenon of color reconnection expresses the possibility
that certain non-leading terms usually ignored in
QCD calculations can drastically
influence the color singlet structure of an event.
Most recent attention to color reconnection has
focused on its implications for the
W boson mass measurement at
LEP-2~\cite{bib-lepww}.
Color reconnection
is an interesting phenomenon in its own right,
however,
as a basic issue of QCD interference and confinement.
In {\epem}$\rightarrow\,$Z$^0\rightarrow\,$$hadrons$ events,
color reconnection
is expected to occasionally yield an event in which 
a pure system of gluons hadronizes in isolation from 
the rest of the event
(see~\cite{bib-lepww,bib-gustafson} and the discussion
below in Sect.~\ref{sec-reconnection}).
Such events are expected to markedly affect the mean properties
of events in which the initial quark and antiquark from
the decay of the Z$^0$ recoil against a gluon jet,
as in our selected
{\epem}$\rightarrow\,${\qq}$\,${\gincl} sample.
Thus,
our gluon jet data can provide a sensitive
test of the color reconnection phenomenon.
In this paper,
we use our {\gincl} data to
perform the most stringent test to date
of the model for color reconnection~\cite{bib-lonnblad}
implemented in the Ariadne
Monte Carlo multihadronic event generator~\cite{bib-ariadne},
version~4.08.

\section{Detector and data sample}
\label{sec-detector}

The OPAL detector is described in
detail elsewhere~\cite{bib-detector,bib-si}.
The tracking system
consists of a silicon microvertex detector,
an inner vertex chamber,
a large volume jet chamber
and specialized chambers at the outer radius of the jet chamber
which improve the measurements in the
$z$-direction.\footnote{Our
coordinate system is defined so that
$z$~is the coordinate parallel to the e$^-$ beam axis,
$r$~is the coordinate normal to the beam axis,
$\phi$~is the azimuthal angle around the beam axis and
$\theta$~is the polar angle \mbox{with respect to~$z$.}}
The tracking system covers the region
$|\cos\theta|$$\,<\,$0.98 and
is enclosed by a solenoidal magnet coil
with an axial field of~0.435~T.
Electromagnetic energy is measured by a
lead-glass calorimeter located outside the magnet coil,
which also covers $|\cos\theta|$$\,<\,$0.98.

The present analysis is based on a sample of
about $3\,708\,000$ hadronic Z$^0$ decay events,
corresponding to our data sample from \mbox{LEP-1}
which includes readout of the silicon strip 
microvertex detector~\cite{bib-si}:
$998\,940$ of these events
were collected in 1991 and~1992 when our 
microvertex detector was instrumented
for readout of the $r$-$\phi$ coordinate only,
while the remainder of the events,
collected from 1993 to 1995,
contain readout
of both the $r$-$\phi$ and $z$ coordinates.
The procedures for identifying
hadronic events are discussed in~\cite{bib-opaltkmh}.
Charged tracks and electromagnetic clusters were
selected for the analysis as follows.
Charged tracks were required to have at least 20 measured
points (of 159 possible) in the jet chamber,
to have a momentum greater than 0.10~GeV/$c$,
to lie in the region $|\cos\theta|$$\,<\,$0.94,
and to point to the origin to within 5 cm in the $r$-$\phi$ plane.
In addition, they were required to yield a $\chi^2$ per
degree-of-freedom of less than~100
for the track fit in the $r$-$\phi$ plane.
Clusters were required to be spread over at least
two lead glass blocks
and to have an energy greater than 0.10~GeV if they
were in the barrel section of the detector ($|\cos\theta|$$\,<\,$0.82)
or greater than 0.30~GeV if they were in
the endcap section (0.82$\,<\,$$|\cos\theta|$$\,<\,$0.98).
Each accepted track and cluster
was considered to be a particle.
Tracks were assigned the pion mass.
Clusters were assigned zero mass since they originate
mostly from photons.
To eliminate residual background and events
in which a significant number of particles was lost
near the beam direction,
the number of accepted charged tracks in each event
was required to be at least five and the 
thrust axis~\cite{bib-thrust} of the event,
calculated using the particles,
was required to satisfy
$|\cos (\theta_{\mathrm{thrust}})|$$\,<\,$0.90,
where $\theta_{\mathrm{thrust}}$ is the
angle between the thrust and beam axes.
The residual background to the sample of hadronic events
from all sources was estimated to be less than~1\%.

\section{Gluon jet selection}
\label{sec-gluon}

For this study,
a gluon jet is defined inclusively
as the particles in an {\epem} event hemisphere
opposite to a hemisphere containing
two identified quark jets,
as stated in the introduction.
To select the {\gincl} gluon jets,
each event is divided into hemispheres using the
plane perpendicular to the thrust axis.
The procedures described below are applied
to each hemisphere separately.
For the purpose of identifying
two quark jets in a single hemisphere,
we employ the {\durham} (``Durham'')
jet algorithm~\cite{bib-durham}.
The results for the gluon jet properties are
almost entirely insensitive to this choice of 
jet algorithm,
as is discussed in~\cite{bib-jwg}
(see also Sect.~\ref{sec-systematic}).
Note that a jet algorithm is used 
only as a selection tool for the {\gincl} jets,
not for the analysis of quark jet properties.
The manner in which quark jets are selected
so as to correspond to the definition employed 
by analytic calculations
is presented in Sect.~\ref{sec-uds}.
The resolution parameter of the jet algorithm
is adjusted to yield 
exactly two reconstructed jets in a hemisphere.
Next, we attempt to reconstruct a displaced
secondary vertex in each of the two jets.
Displaced secondary vertices are associated with 
heavy quark decay,
especially that of the b quark.
At LEP, b quarks are produced almost exclusively
at the electroweak vertex\footnote{About 22\% of
hadronic Z$^0$ events contain a b$\overline{\mathrm{b}}$
quark pair from the electroweak decay 
of the Z$^0$~\cite{bib-lepew}
compared to only about 0.2\% with
a b$\overline{\mathrm{b}}$ pair from
gluon splitting~\cite{bib-delphibb}.}:
thus a jet containing a b hadron is almost always a 
quark jet.

To reconstruct secondary vertices in jets,
we employ the method described in~\cite{bib-qg95a}.
Briefly,
charged tracks are selected for the secondary 
vertex reconstruction procedure if they have coordinate
information from at least one of the two silicon detector layers,
if their momentum is larger than 0.5~GeV/$c$,
and if their distance of closest approach
to the primary event vertex~\cite{bib-qg95a} 
is less than 0.3~cm.
Additionally,
we require the maximum uncertainty on the
distance of closest approach to be~0.1~cm.
For the 1991-92 data
(with only \mbox{$r$-$\phi$} coordinate readout
of the microvertex detector),
the distance of closest approach,
and the distances $b$ and $L$
discussed below,
are determined in the \mbox{$r$-$\phi$} plane.
For the 1993-95 data
(with \mbox{$r$-$\phi$} and $z$ coordinate
readout of the microvertex detector),
these distances are determined in three dimensions.
A secondary vertex is required to contain at least three tracks
which satisfy the above criteria.
For the 1991-92 data,
at least two of these tracks are required to
satisfy $b/\sigma_b$$\,>\,$2.5,
where $b$ is the signed impact parameter value
of a track with respect to the primary event vertex
and $\sigma_b$ is the uncertainty associated with~$b$.
For the 1993-95 data,
only one track in the secondary vertex
is required to have $b/\sigma_b$$\,>\,$2.5.
For jets with such a secondary vertex,
the signed decay length, $L$,
is calculated with respect to the primary vertex,
along with its error,~$\sigma_L$.
The sign of $L$ is determined by summing the three momenta
of the tracks fitted to the secondary vertex;
$L$$\,>\,$0 if the secondary vertex is displaced from the primary
vertex in the same hemisphere as this momentum sum,
and $L$$\,<\,$0 otherwise.
The sign of $b$ is determined in an analogous manner.
More details concerning the determination of
$L$ and $b$ are given in~\cite{bib-qg95a}.
To be tagged as a quark jet,
a jet is required to have a visible energy of at least 10~GeV
and a successfully reconstructed secondary vertex
with $L/\sigma_L$$\,>\,$3.5 for the 1991-92 data or
$L/\sigma_L$$\,>\,$5.0 for the 1993-95 data.
The visible energy of a jet is defined by the sum of
the energies of the particles assigned to the jet.
We refer to a hemisphere with two tagged jets
as a tagged hemisphere.

We next examine the angles that the two jets
in a tagged hemisphere
make with respect to the thrust axis and to each other.
If the two jets are close together,
or if one of the two jets is much more energetic
than the other,
it is very likely that one of the two jets is
a gluon jet due to the strong kinematic similarity
to an event with gluon radiation from a quark.
To reduce this background,
we require the angle between each jet and the
thrust axis to exceed~15$^\circ$ and the angle
between the two jets to exceed~70$^\circ$.
We further require the two jets to lie no
more than 70$^\circ$ from the thrust axis
to eliminate jets near the hemisphere boundary.
These angular restrictions on the quark jet directions
do not affect the good correspondence between
{\gincl} jets from {\epem} annihilations and 
hemispheres of {\gluglu} events from a point source,
as is demonstrated below in Sect.~\ref{sec-ggcompare}.
Last, we eliminate events with three tagged jets,
i.e. events in which both jets in one hemisphere and one of the
two jets in the other hemisphere have been tagged as b jets
(about 4\% of the events after the other cuts
have been applied),
because Monte Carlo study shows them 
to be mostly background.\footnote{Such events can arise
from gluon splitting to a b$\overline{\mathrm{b}}$ pair;
although rare in inclusive Z$^0$ decays,
our analysis preferentially selects such events.}
There are no events in which both jets in
both hemispheres are tagged.
In total,
439 events are selected for the gluon jet
{\gincl} sample:
87 from the 1991-92 data and 352 from
the 1993-95 data.
The mean angle between the two tagged quark jets
in the final {\gincl} jet sample is 91.5$^\circ$
with a standard deviation of~12.8$^\circ$.

The purity of this sample is estimated using the
Jetset Monte Carlo multihadronic event generator~\cite{bib-jetset}
including detector simulation~\cite{bib-gopal} 
and the same analysis
procedures as are applied to the data.
For the simulation of the 1991-92 data,
we use a combination of events generated using
version~7.3 of the program with the parameter values
given in~\cite{bib-qg93}
and of events generated using
version~7.4 of the program with the parameter values
given in~\cite{bib-qg95b}.
The initial Monte Carlo samples have about 
$3\,000\,000$ events for version~7.3 and
$1\,000\,000$ events for version~7.4.
The two Jetset versions yield results which are consistent
with each other to within the statistical uncertainties
and so we combine them.
For the simulation of the 1993-95 data,
we use a sample of about $6\,000\,000$ events
generated using version~7.4 with the parameters
given in~\cite{bib-qg95b}.
The hadron level Monte Carlo jets are examined
to determine whether they are associated with an
underlying quark or antiquark jet.
To perform this association,
the Monte Carlo events are also examined at the parton level.
We determine the directions of the primary
quark and antiquark from the Z$^0$ 
decay after the parton shower has terminated.
The hadron jet closest to 
the direction of an evolved primary quark or antiquark
is considered to be a quark jet.
The distinct hadron jet closest to the evolved primary
quark or antiquark not associated with this first
hadron jet is considered to be the other quark jet.
An event in which one of the two tagged jets is
{\it not} identified as a quark jet
is deemed to be a background event.
Using this algorithm,
we estimate the purity of the {\gincl} sample
to be $(78.8\pm 2.4\,\mathrm{(stat.)})$\% for the 1991-92 data
and $(82.9\pm 1.4\,\mathrm{(stat.)})$\% for the 1993-95 data.
The estimated purity of the combined 1991-1995 sample is
$(81.9\pm 1.2\,\mathrm{(stat.)})$\%.
The background events mostly arise when two tracks
from a long lived particle such as a 
K$^0_{\mathrm{S}}$ or $\Lambda$ are combined
with a third track to define a secondary vertex in a gluon jet,
or else from events in which a gluon decays into
a b$\overline{\mathrm{b}}$ pair.
About 94\% of the events in the 
{\gincl} sample are predicted to be b events.
This reliance on b events is not expected to affect our results
since the properties of hard, acollinear gluon jets
do not depend on the event flavor according to QCD.
More details are given in~\cite{bib-opalqg96}.

The {\gincl} tag rates,
defined by the ratio of the number of {\gincl} jets
to the number of events in the initial
inclusive multihadronic event samples,
are $(8.71\pm 0.94\,\mathrm{(stat.)})\times 10^{-5}$
for the data and
$(7.20\pm 0.42\,\mathrm{(stat.)})\times 10^{-5}$
for the Monte Carlo for the 1991-92 analysis,
and $(1.30\pm 0.07\,\mathrm{(stat.)})\times 10^{-4}$
for the data and
$(1.25\pm 0.05\,\mathrm{(stat.)})\times 10^{-4}$
for the Monte Carlo for the 1993-95 analysis.
Thus the Monte Carlo reproduces the measured tag rates well.
The tag rate for the 1993-95 data
is substantially larger than that for the 1991-92 data
as a consequence of the addition of $z$~coordinate readout
from the silicon microvertex detector.

The energy of the {\gincl} jet is determined
by imposing overall energy-momentum conservation
on the system of three jets comprised of
the {\gincl} jet
and the two jets in the tagged hemisphere.
A direction is determined for the {\gincl} jet by
summing the momenta of the particles in its hemisphere.
The jet directions are used in conjunction with
the jet velocities to calculate the jet energies,
assuming massive kinematics.\footnote{
For a system of three jets labelled 1, 2 and 3,
the energy-momentum constraints
$\displaystyle{\Sigma_{i=1}^{3}E_{\,{\mathrm{jet}}\,i}}$={\ecm}
and
$\displaystyle{\Sigma_{i=1}^{3}\vec{P}_{\,{\mathrm{jet}}\,i}}$=0
are solved for $E_{\,{\mathrm{jet}}\,i}$,
where 
$\displaystyle{\vec{P}_{\,{\mathrm{jet}}\,i}}$=
$\displaystyle{\vec{\beta}_i
E_{\,{\mathrm{jet}}\,i}}$
is the momentum of jet~$i$,
with its velocity $\vec{\beta}_i$
given by its visible 3-momentum divided by its visible 
energy.
The solution is
$E_{\,{\mathrm{jet}}\,i}$=
{\ecm}$\beta_j\beta_k\sin\theta_i
/(\beta_1\beta_2\sin\theta_3+
\beta_1\beta_3\sin\theta_2+
\beta_2\beta_3\sin\theta_1)$
where $\theta_1$, $\theta_2$ and $\theta_3$
are the angles between the jets
with $\theta_i$ opposite to jet~$i$,
and where $(i,j,k)$=$(1,2,3)$, $(2,3,1)$ or $(3,1,2)$.
}
We obtain {\egincl}$\,$=$\,$$\,40.1\pm 0.2\,\mathrm{(stat.)}$~GeV.
This value includes a multiplicative correction of
1.03 to account for the effects of detector response
and initial-state photon radiation.
The correction factor is obtained using Monte Carlo
predictions with and without simulation of the
detector as is described in~\cite{bib-opalqg96}
(see also Sect.~\ref{sec-corrections}).
The corrected mean visible energy of the {\gincl} jets
is $40.8\pm 0.4\,\mathrm{(stat.)}$~GeV.
The correction procedure accounts for any possible
double counting of particle energy in the determination
of the jet's visible energy.
The difference between the mean calculated
and visible jet energies is used to define a
systematic uncertainty.
The mean energy of the gluon jets in our study is
therefore
{\egincl}$\,$=$\,$$\,40.1\pm 0.2\,\mathrm{(stat.)}
\pm 0.7\,\mathrm{(syst.)}$~GeV.

\section {Light quark jet selection}
\label{sec-uds}

To select quark jets in
a manner which corresponds to analytic calculations,
we define quark jets inclusively
as the particles in hemispheres
of light (uds) flavored 
{\epem}$\rightarrow\,$Z$^0\rightarrow\,$$hadrons$
events.
Note that these are not the same as the quark jets
discussed in the previous section
(defined using the {\durham} jet algorithm),
which are
used only as a tool to identfy {\gincl} gluon jets.
Due to the relatively large efficiency of the
uds jet selection procedure (see below),
it is not necessary to employ the entire sample
of about $3\,708\,000$ events
mentioned in Sect.~\ref{sec-detector} 
for the uds jet analysis.
Instead,
we base this analysis on an initial sample of
$222\,921$ hadronic annihilation
events with c.m. energies 
within~100~MeV of the Z$^0$ peak.
In addition to the selection criteria described in
Sect.~\ref{sec-detector},
we require the angle $\theta_{\mathrm{thrust}}$
between the thrust and beam axes to satisfy
$|\cos (\theta_{\mathrm{thrust}})|$$\,<\,$0.70 for this analysis,
to contain the events well within
the geometric acceptance of the silicon microvertex detector.

The uds jet tagging is based on 
the signed impact parameter values of charged tracks
with respect to the primary event vertex, $b$,
since the distribution of this variable
is strongly skewed toward positive values 
for c and b events but not for uds events.
Charged tracks are selected for the uds tagging procedure
if they have $r$-$\phi$ coordinate information 
from at least one silicon detector layer,
a momentum of 0.5~GeV/$c$ or larger,
and a maximum distance of closest approach
to the primary event vertex in the $r$-$\phi$ plane of 0.3~cm
with a maximum uncertainty on this quantity of~0.1~cm.
If no track in an event satisfies these requirements
(0.003\% of the events), the event is eliminated.
The number of tracks
which meet these requirements and which have 
$b/\sigma_b$$\,>\,$2.5
in the $r$-$\phi$ plane is determined.
An event is tagged as containing a uds jet if 
this number is zero.
In total, $53\,552$ events are tagged.
Both hemispheres of a tagged event are identified as
uds jet hemispheres and are used in the
subsequent analysis:
thus, there are $107\,104$ uds jets in our study.
The estimated uds purity of this sample,
obtained by treating Jetset events with detector simulation
in the same manner as the data, 
is~$(86.4\pm 0.3\,\mathrm{(stat.)})$\%.
The Monte Carlo predicts that 86\% of the background
events are c events and that 14\% are b events.
The uds jet tag rate,
defined by the ratio of the number of identified
uds jets to the number of events in the initial
inclusive multihadronic event sample, is
$0.480\pm 0.002\,\mathrm{(stat.)}$
for the data and 
$0.487\pm 0.001\,\mathrm{(stat.)}$
for the Monte Carlo:
thus the measured and simulated tag rates
agree to better than~1\%.
The energy of the uds jets is given by
the beam energy, 45.6~GeV,
with essentially no uncertainty.

\section{Experimental distributions}
\label{sec-distr}

In this study, we examine the distributions of rapidity,
scaled energy, the logarithm of the momentum,
transverse momentum with respect to the jet axis,
and multiplicity in restricted rapidity intervals,
of charged particles in the {\gincl} and uds jets.
All these variables are commonly used
to characterize the energy and multiplicity structure of jets.

Rapidity, $y$, is defined by
\begin{equation}
y=\frac{1}{2}\ln\left(\frac{E+\vec{p}\cdot\hat{r}}
     {E-\vec{p}\cdot\hat{r}}\right)
\end{equation}
with $E$ and $\vec{p}$ the energy and momentum of
a particle and $\hat{r}$ the axis with respect to which
rapidity is calculated.
We choose $\hat{r}$ to be the sphericity 
axis~\cite{bib-sph} calculated using the charged and neutral
particles in the {\gincl} or uds jets.
We do not use the thrust axis to calculate rapidity,
contrary to common usage,
because the thrust axis is used to determine the hemisphere
boundaries of the {\gincl} and uds jets and we wish to
reduce the correlation between the event 
selection and the jet analysis.

The scaled energy of a particle, $x_{E}$, is given by
\begin{equation}
  {x_{E}}=\frac{E}{E_{\,\mathrm{jet}}}
\end{equation}
with {\ejet} determined as explained in
Sects.~\ref{sec-gluon} and~\ref{sec-uds}.
The distribution of $x_{E}$ is commonly referred to
as the fragmentation function.

We also study the distributions of \mbox{$\ln\,(p)$},
with $p$ the particle momentum,
and of $\pperp$,
the transverse momentum of particles
with respect to the jet axis,
as stated above.
The jet axis for the $\pperp$ calculation is
defined by summing the momenta of the particles
in the {\gincl} or uds jets.
Besides the inclusive $\pperp$ distribution,
we examine the $\pperp$ distribution of soft particles,
defined as particles
with momenta below 4~GeV/$c$:
we refer to this distribution as~$\ptsoft$.
The motivation for including this last variable
in our study is presented in Sect.~\ref{sec-softpt}.

Last,
we study the distribution of charged particle
multiplicity in restricted intervals of phase space,
specifically for {\mysph}$\,\leq\,$2 and \mbox{{\mysph}$\,\leq\,$1.}
This complements our study of the charged
particle multiplicity distributions of
{\gincl} and uds jets in full phase space,
presented in~\cite{bib-opalqg97}.
The distribution of charged multiplicity in restricted
regions of phase space is more sensitive to the dynamics
of multihadron production
than the distribution in full phase space
because it is less affected by the constraints of overall 
charge and energy-momentum
conservation.

With the exception of multiplicity,
the distributions in this paper are normalized by
the number of events in the sample.
The multiplicity distributions are normalized
to have unit~area.

\section{Monte Carlo comparison of {\boldmath{\gincl}} and
{\boldmath{\gluglu}} jets}
\label{sec-ggcompare}

Our analysis of gluon jets is based on the premise
that {\gincl} jets from {\epem}
annihilations are equivalent to
hemispheres of {\gluglu} events produced from
a color singlet point source,
with the hemispheres defined by the plane
perpendicular to the principal event axis.
Although high energy {\gluglu} events are not
available experimentally,
they may be generated using a 
QCD Monte Carlo event generator.
The viability of our premise can be tested
by comparing the Monte Carlo predictions for {\gluglu} 
event hemispheres and {\gincl} jets.
Such a comparison was
presented in~\cite{bib-opalqg97}
for the charged particle multiplicity distribution
in full phase space (see also~\cite{bib-jwg}).
In this section we extend this comparison to the
distributions studied here.

\begin{figure}[ht]
\epsfxsize=17cm
\epsffile[65   100 595 700]{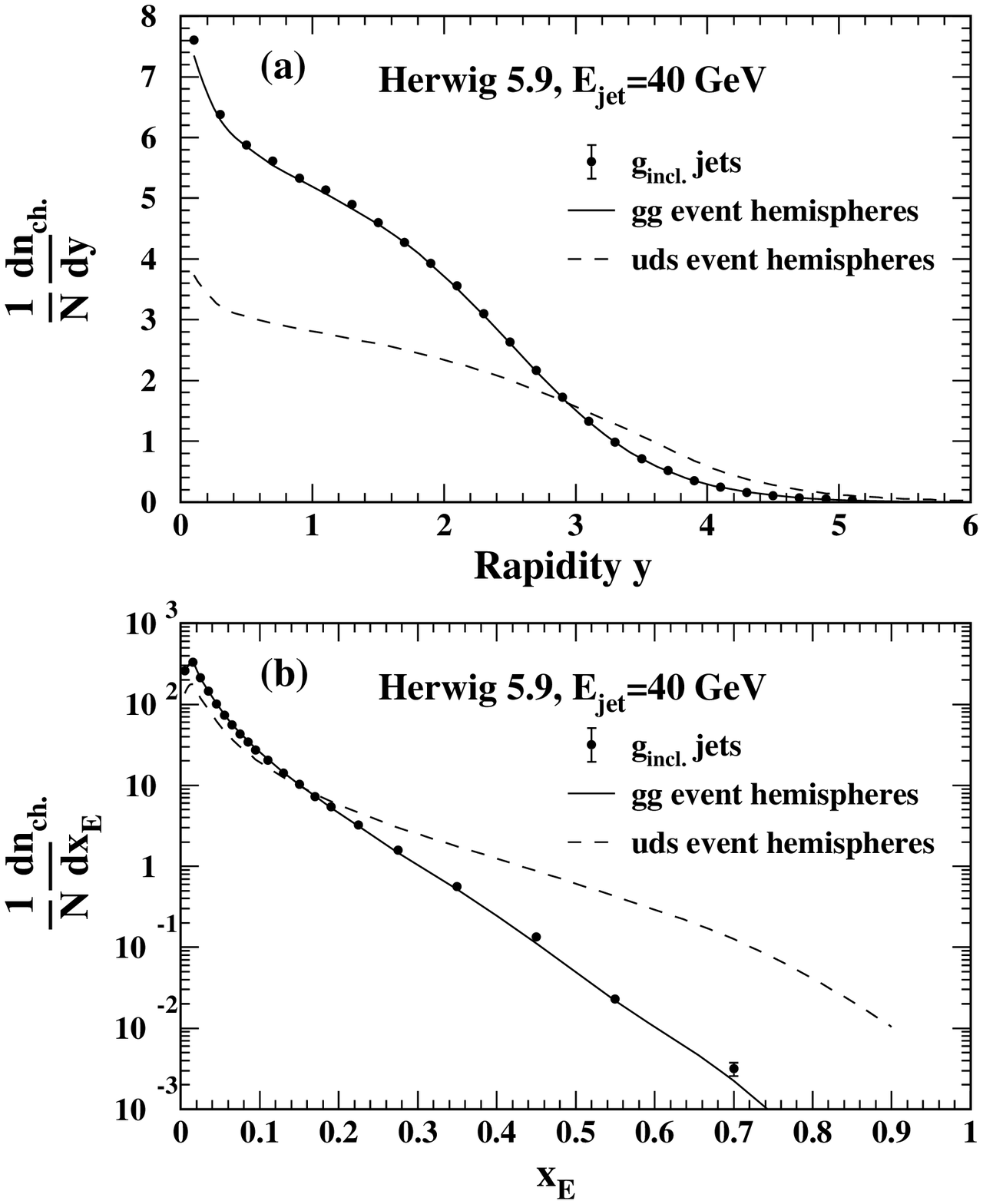}
\caption{
The prediction of the Herwig parton shower
Monte Carlo event generator for the charged particle
(a)~rapidity 
and (b)~{\xee}$\,$=$\,$$E/E_{\,\mathrm{jet}}$
distributions of {\gincl} gluon jets 
from {\epem} annihilations,
in comparison to the Herwig predictions for
{\gluglu} and uds event hemispheres.
The jet energies are 40~GeV,
corresponding to a c.m. energy
of 91.2~GeV for the generation of the
{\epem}$\rightarrow\,${\qq}$\,${\gincl} events.
}
   \label{fig-hwigpage1}
\end{figure}

The solid points in Fig.~\ref{fig-hwigpage1}a
show the prediction of the Herwig Monte Carlo
multihadronic event generator~\cite{bib-herwig}, version~5.9,
for the charged particle rapidity distribution
of {\gincl} jets in {\epem}$\rightarrow\,${\qq}$\,${\gincl} events.
The uncertainties are statistical
(these are too small to be visible).
The parameter set we use is the same as that
given in~\cite{bib-qg95b} for Herwig, version~5.8,
except that the value of the cluster mass cutoff 
CLMAX has been increased from 3.40~GeV/$c^2$
to 3.75~GeV/$c^2$
to improve the model's description of
the mean charged particle multiplicity~{\mnch}
in inclusive hadronic Z$^0$ 
\mbox{decays~\cite{bib-opaltransverse}-\cite{bib-alephes}.}
The {\epem}$\rightarrow\,${\qq}$\,${\gincl} events
were generated using a center-of-mass (c.m.) energy, 
{\ecm}, of 91.2~GeV to correspond to the data.
The {\gincl} identification
was performed using the same procedure as is
described for the data in Sect.~\ref{sec-gluon},
except that the two quark jets against which
the {\gincl} jet recoils were identified
using parton level Monte Carlo information
as described in Sect.~\ref{sec-gluon}
rather than using displaced secondary vertices.
In particular,
the angular restrictions on the directions of the quark jets with
respect to the thrust axis and to each other have been applied.
The resulting mean energy of the Monte Carlo
{\gincl} jets is~40.0~GeV
with a negligible statistical uncertainty.

The solid curve in
Fig.~\ref{fig-hwigpage1}a shows the
rapidity distribution predicted by
Herwig for {\gluglu} event hemispheres.
The {\gluglu} events were generated using a c.m. energy
of 80.0~GeV so that
the hemisphere energies are the same as for the {\gincl} jets.
It is seen that the results for {\gincl} jets
and {\gluglu} event hemispheres are 
almost indistinguishable.
This establishes the validity of our technique
to identify gluon jets in a manner which corresponds
to point source production from a
color singlet~\cite{bib-jwg}.
Similar agreement between the predicted rapidity
distributions of {\gincl} jets and {\gluglu} event hemispheres 
is obtained if Jetset is used to generate the
samples rather than Herwig,
or if the JADE-E0~\cite{bib-jade}
or cone~\cite{bib-opalcone}
jet finder is used to identify the quark jets
for the {\gincl} jet selection
rather than the {\durham} jet finder.
This emphasizes the independence of our results from
the choice of a jet finding algorithm.
For purposes of comparison,
the dashed curve in
Fig.~\ref{fig-hwigpage1}a
shows the prediction of Herwig for uds event hemispheres,
generated using the same c.m. energy
as is used to generate the {\gluglu} event sample.

\begin{figure}[ht]
\epsfxsize=17cm
\epsffile[65   100 595 700]{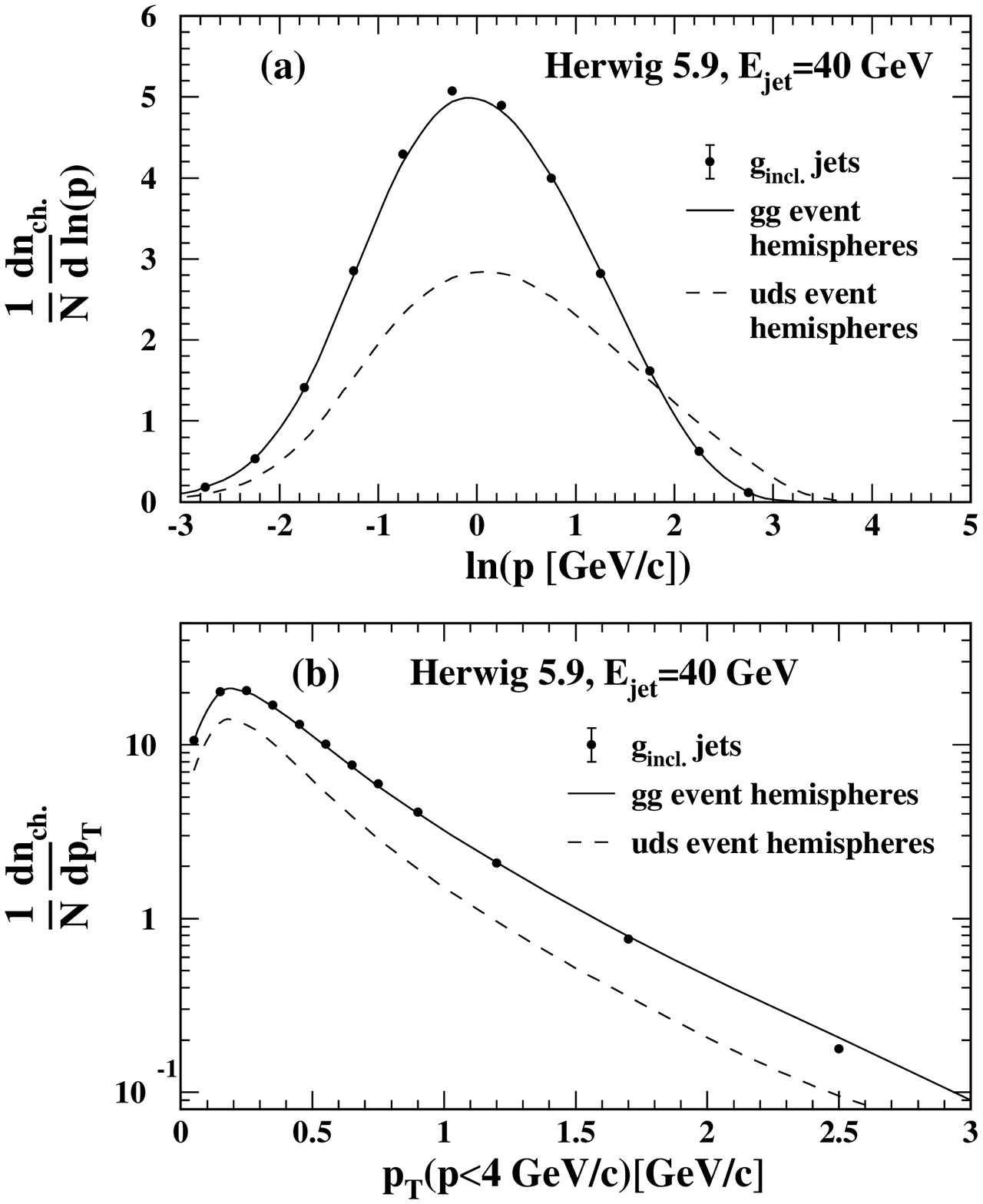}
\caption{
The prediction of the Herwig parton shower
Monte Carlo event generator for the charged particle
(a)~$\ln\,(p)$
and (b)~$\ptsoft$ distributions
of {\gincl} gluon jets 
from {\epem} annihilations,
in comparison to the Herwig predictions for
{\gluglu} and uds event hemispheres.
The jet energies are 40~GeV,
corresponding to a c.m. energy
of 91.2~GeV for the generation of the
{\epem}$\rightarrow\,${\qq}$\,${\gincl} events.
}
   \label{fig-hwigpage2}
\end{figure}

\begin{figure}[ht]
\epsfxsize=17cm
\epsffile[65   100 595 700]{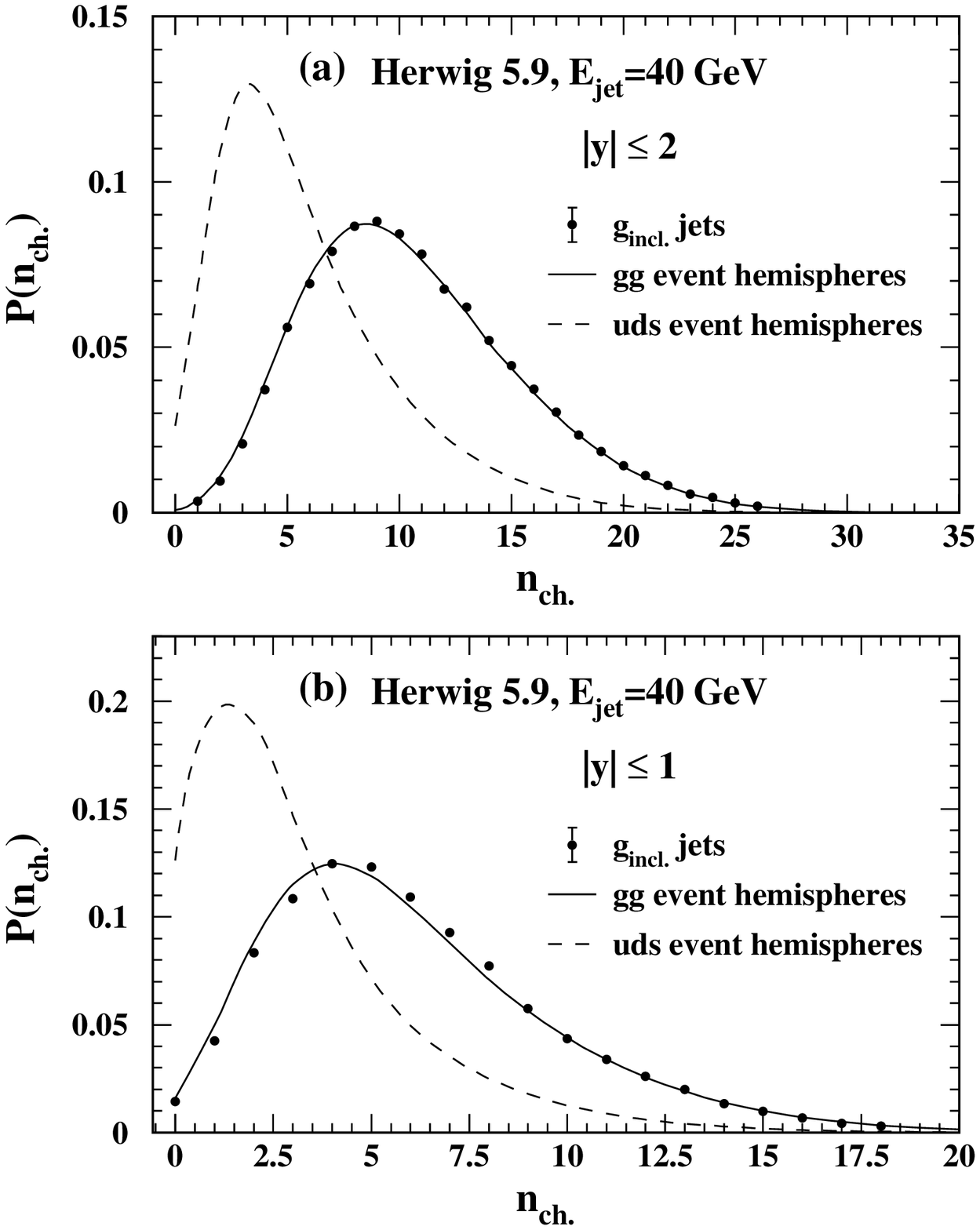}
\caption{
The prediction of the Herwig parton shower
Monte Carlo event generator for the charged particle
multiplicity distributions of {\gincl} gluon jets 
from {\epem} annihilations,
in the rapidity intervals (a)~{\mysph}$\,\leq\,$2 
and (b)~\mbox{{\mysph}$\,\leq\,$1},
in comparison to the Herwig predictions for
{\gluglu} and uds event hemispheres.
The jet energies are 40~GeV,
corresponding to a c.m. energy
of 91.2~GeV for the generation of the
{\epem}$\rightarrow\,${\qq}$\,${\gincl} events.
}
   \label{fig-hwigpage3}
\end{figure}

The corresponding Monte Carlo comparison of 
the properties of {\gincl} jets
with those of {\gluglu} and uds event hemispheres
is shown in Fig.~\ref{fig-hwigpage1}b
for the {\xee} distribution,
in Fig.~\ref{fig-hwigpage2} for the $\ln\,(p)$
and $\ptsoft$ distributions,
and in Fig.~\ref{fig-hwigpage3} for the charged particle
multiplicity distributions with 
\mbox{{\mysph}$\,\leq\,$2} and
\mbox{{\mysph}$\,\leq\,$1}.
The results for the inclusive $\pperp$ distribution are
qualitatively similar to those shown in 
Fig.~\ref{fig-hwigpage2}b
for the $\ptsoft$ distribution and so we do not
show them in addition.
The results for {\gincl} jets are seen to reproduce
those of {\gluglu} event hemispheres with good accuracy.
In Fig.~\ref{fig-hwigpage3},
a small shift is observable between the multiplicity
distributions of {\gincl} and {\gluglu} event hemispheres
at intermediate values of multiplicity,
with the {\gincl} jets exhibiting slightly larger multiplicity values.
This shift is more pronounced for
smaller rapidity intervals,
i.e., it is more pronounced in
Fig.~\ref{fig-hwigpage3}b than in Fig.~\ref{fig-hwigpage3}a
(no such shift is visible for the charged particle
multiplicity distribution in full phase space,
see Fig.~1 in~\cite{bib-opalqg97}).
This difference between {\gincl} jets and {\gluglu}
event hemispheres
is negligible compared to the experimental
statistical uncertainties (Sect.~\ref{sec-results})
or to the difference between the uds and gluon jets
and so we ignore it.
Thus {\gincl} jets from {\epem} events
have almost identical properties to gluon jets
in {\gluglu} events produced from a color singlet point source,
as stipulated in the introduction.

\section{Corrections}
\label{sec-corrections}

To correct the data for detector response
and initial-state photon radiation,
we generate events with the Jetset Monte Carlo
and compare their properties with and without
simulation of the detector and with and without
initial-state radiation.
The data are corrected to the hadron level.
The hadron level does not include detector simulation
or initial-state radiation and treats
all charged and neutral particles with lifetimes greater
than \mbox{$3\times 10^{-10}$~s} as stable:
hence charged particles from the decays of K$_{\mathrm S}^0$
and weakly decaying hyperons are included in the
corrected distributions.
The corrections account not only for 
detector response and initial-state radiation but also 
for the background to the {\gincl} and uds events.
There is good agreement between the data and
Monte Carlo
for Monte Carlo samples which include background,
initial-state radiation,
detector simulation, and the same analysis procedures
as are applied to the data.
Furthermore, 
systematic shifts observed between Jetset and the data
before the corrections are applied
are also observed after corrections.
For example,
the mean charged particle multiplicity of {\gincl} jets
in full phase space is 1\% larger in Jetset than in the
data before corrections and 3\% larger after corrections.
Thus the corrections do not introduce a significant
bias towards the predictions of Jetset.

Since the analysis of {\gincl} jets is somewhat different
for the 1991-92 and 1993-95 data samples
(Sect.~\ref{sec-gluon}),
separate corrections are determined for them.
The corrected results from the two samples are
consistent with each other to within their statistical uncertainties.
The final gluon jet results are obtained by forming the weighted
mean of the corrected 1991-92 and 1993-95 results.

The distributions of {\ysph}, {\xee}, $\ln\,(p)$,
$\pperp$ and $\ptsoft$
are corrected
using bin-by-bin multiplicative factors
which are determined as described in~\cite{bib-opalqg96}.
The corrections are typically in the
range between 0.90 and 1.15 for both
the {\gincl} and uds jets.
For the distributions of charged particle multiplicity
with {\mysph}$\,\leq\,$2 and \mbox{{\mysph}$\,\leq\,$1},
we do not utilize simple bin-by-bin corrections
since the Monte Carlo predicts considerable migration 
between multiplicity bins as a consequence of
the detector response.
Instead, the data are corrected in a two stage
process using the method described in~\cite{bib-opalqg97}.
In the first stage,
the data are corrected for experimental acceptance, resolution, 
and secondary electromagnetic and hadronic interactions
using an unfolding matrix~\cite{bib-opalnch}.
In the second stage,
the data are corrected for background,
geometric event acceptance,
and the effects of initial-state radiation
using bin-by-bin factors.
More details are given in~\cite{bib-opalqg97}.
As an indication of the overall size of the corrections,
Jetset predicts the mean multiplicity value
of {\gincl} jets for {\mysph}$\,\leq\,$2 ({\mysph}$\,\leq\,$1)
to be 11\% (13\%)
larger at the hadron level
than it is at the level which includes
background, initial-state radiation,
detector simulation,
and the experimental selection criteria.
The corresponding difference for uds jets is~0\% (+1\%).

\section{Systematic uncertainties}
\label{sec-systematic}

To evaluate systematic uncertainties,
the analysis was repeated with the
following changes relative to the standard analysis.
There were no significant changes in the number
of selected events or in the estimated purities 
of the gluon and quark jet samples
compared to the standard results 
(439~{\gincl} jets with 81.9\% purity
and $107\,104$~uds jets with 86.4\% purity)
unless otherwise noted.
\begin{enumerate}
\item  Charged tracks alone were used for the data and
  for the Monte Carlo samples with detector simulation,
  rather than charged tracks plus electromagnetic clusters;
  the number of selected {\gincl} jets decreased to~327.
  As an additional check on the track selection,
  the minimum momentum of charged tracks
  was increased from
  from 0.10~GeV/$c$ to 0.25~GeV/$c$.
\item  Herwig was used to determine the corrections
  for background, detector response and initial-state radiation,
  rather than Jetset.
\item  The gluon jet selection was performed using the
  JADE-E0 and cone
  jet finders to define the tagged quark jets,
  rather than the {\durham} jet finder;
  for the analysis based on the cone jet finder,
  the number of {\gincl} jets dropped to~346
  while their estimated purity decreased to~73.2\%.
\item  The geometric conditions for the
  gluon jet selection were varied, 
  first by requiring the angle between the two
  jets in the tagged hemisphere to exceed 50$^\circ$,
  rather than~70$^\circ$,
  and second by requiring the two tagged quark jets
  to lie within 65$^\circ$ of the thrust axis,
  rather than~70$^\circ$.
  For the first of these conditions,
  the {\gincl} sample increased to 583~jets
  while its estimated purity decreased to~73.5\%;
  for the second of these conditions,
  the {\gincl} sample decreased to 383~jets.
\item Secondary vertices used to tag quark
  jets for the {\gincl} identification were required to
  have decay lengths which satisfied $L/\sigma_L$$\,>\,$5.0,
  rather than $L/\sigma_L$$\,>\,$3.5,
  for the 1991-92 data,
  and $L/\sigma_L$$\,>\,$7.0,
  rather than $L/\sigma_L$$\,>\,$5.0,
  for the 1993-95 data;
  the {\gincl} sample decreased to 268 jets
  while its estimated purity increased to~84.9\%.
\item  Tracks selected for the uds tagging procedure
  were required to have a signed impact parameter
  which satisfied $b/\sigma_b$$\,>\,$1.5,
  rather than $b/\sigma_b$$\,>\,$2.5;
  the uds sample decreased to $49\,396$ jets
  while its estimated purity increased to~89.9\%.
  As an additional check on the track selection
  for the uds tagging procedure,
  tracks used for this procedure were 
  required to satisfy the following criteria:
  (i)~the maximum 
  distance of closest approach of the track to the primary 
  event vertex in the $r$-$\phi$ plane was 5.0~cm, 
  rather than 0.3~cm,
 (ii)~no requirement was placed on the uncertainty of
  the distance of closest approach of the track to the
  event vertex,
  rather than requiring this uncertainty to be less
  than~0.1~cm in the $r$-$\phi$ plane,
  and (iii)~the minimum momentum was 0.1~GeV/$c$
  rather than 0.5~GeV/$c$;
  the uds sample decreased to $70\,826$ jets.
\item  For the ratios of the gluon to quark jets results
  (see Sect.~\ref{sec-results}),
  the energy to which the quark jet results were
  corrected was varied by the total uncertainty
  of the {\gincl} jet energy (Sect.~\ref{sec-gluon});
  also, for these same quantities,
  the correction factors to account for the difference
  between the uds and {\gincl} jet energies 
  (Sect.~\ref{sec-results})
  were varied by their uncertainties.
\end{enumerate}
The differences between the standard results
and those found using each of these conditions
were used to define symmetric systematic uncertainties.
For items~1, 3, 4, 6 and~7,
the larger of the two described differences
with respect to the standard result
was assigned as the systematic uncertainty.
For item~2,
the difference with respect to the standard result
was multiplied by 2/$\sqrt{12}$ for
the following reason.
Jetset describes the basic experimental 
distributions in this paper very well
(as already stated in Sect.~\ref{sec-corrections})
and in this sense its results represent a
central value compared to the results
of other Monte Carlo simulations.
In contrast,
Herwig disagrees 
with the data for some basic distributions,
especially for uds jets.\footnote{For example,
the mean charged particle multiplicity
of uds jet hemispheres is measured to be
$10.10\pm0.18\,$(stat.+syst.)~\cite{bib-opalqg97},
compared to predictions of
9.67 and 10.03 for our tuned
versions of Herwig and Jetset, respectively:
this represents a difference of
2.4~standard deviations for Herwig
but of only 0.4~standard deviations for Jetset.
The low value of the Herwig prediction for multiplicity
in uds jets is reflected in some of the basic distributions
of our study,
such as the uds jet rapidity distribution
shown in Fig.~\ref{fig-rapidity}a (Sect.~\ref{sec-results}).}
Thus, the results of 
Herwig represent an extreme choice 
for this analysis.
The factor of 2/$\sqrt{12}$ converts the
difference between an extreme and central value into
a dispersion,
i.e. into a more realistic estimate of the 
uncertainty related to the model dependence
of the corrections for detector response.

The systematic uncertainty evaluated for each bin 
of the differential distributions
(Figs.~\ref{fig-rapidity}--\ref{fig-nchy},
see Sect.~\ref{sec-results})
was averaged with the results from its two neighbors
to reduce the effect of bin-to-bin fluctuations
(the single neighbor was used for 
bins on the endpoints of the distributions).
The uncertainties were added in quadrature
to define the total systematic uncertainty.
The largest systematic terms for the {\gincl} jet measurements 
generally arose about equally from items~1, 3, 4 and~5
in the above list.
The largest systematic terms for the uds jet measurements
generally arose about equally from items~2 and~6.
For the ratios of the gluon to quark jet measurements,
the largest systematic terms
generally arose from items~1, 3 and~5.

As an additional systematic check,
events selected for the {\gincl} analysis were required
to have c.m. energies within 100~MeV of the Z$^0$ peak.
This restriction eliminated 12\% of the events from the
{\gincl} sample and resulted in insignificant changes to
the measured {\gincl} jet properties.

\section{Results}
\label{sec-results}

The corrected distributions of
{\ysph}, {\xee}, \mbox{$\ln\,(p)$}, $\pperp$ and~$\ptsoft$
are shown in the top portions of
Figs.~\ref{fig-rapidity}-\ref{fig-ptsoft}.
The corresponding results for charged particle
multiplicity with
{\mysph}$\,\leq\,$2 and \mbox{{\mysph}$\,\leq\,$1}
are shown in Fig.~\ref{fig-nchy}.
Numerical values for these data
are provided in Tables~\ref{tab-rapidity}-\ref{tab-nchyltone}.
The vertical lines on the data points
show the total uncertainties,
with statistical and systematic terms added in quadrature.
The small horizontal lines indicate the size of the
experimental statistical uncertainties.
Our results for the charged particle uds fragmentation function
(Fig.~\ref{fig-xee} and Table~\ref{tab-xee})
are consistent with those presented in~\cite{bib-opalhapke}.
The matrix correction technique employed for the
multiplicity distributions in Fig.~\ref{fig-nchy}
introduces correlations between bins.
These correlations are generally strong between
a bin and its nearest one or two neighbors on either side
but can extend with smaller strength to
four or five bins away.
The correlations smooth out
bin-to-bin statistical fluctuations.
This effect is particularly noticeable for the gluon jet
distributions in Fig.~\ref{fig-nchy}
because of the relatively small number of
events in the {\gincl} jet sample.

For the {\ysph}, {\xee}, \mbox{$\ln\,(p)$}, $\pperp$ 
and~$\ptsoft$ distributions,
we also determine ratios between
the gluon and quark jet measurements since common
systematic uncertainties will partially cancel.
Before forming these ratios,
we account for the different
energies of the two samples:
the gluon jets have a mean energy of 40.1~GeV whereas
the uds jets have a mean energy of 45.6~GeV.
To correct the quark jets for this difference in energy,
we employ multiplicative factors determined bin-by-bin
using Jetset.
As a systematic check,
we also determine the corrections predicted by Herwig.
Since the energy difference between the {\gincl} and uds jets
is only 5.5~GeV,
these corrections are small:
they typically lie between 0.95 and~1.01.
The bottom portions of
Figs.~\ref{fig-rapidity}-\ref{fig-ptsoft}
show the ratios of the gluon to quark jet results,
corresponding to jet energies of~40.1~GeV.
Numerical values for these ratio measurements
are included in Tables~\ref{tab-rapidity}-\ref{tab-ptsoft}.

\subsection{Mean multiplicity ratio at small rapidities}
\label{sec-nchrap}

A striking feature of our results
is the nearly factor of two difference between the
mean multiplicities of gluon and quark jets
at small rapidities and energies
(see~Figs.~\ref{fig-rapidity}b and~\ref{fig-xee}b).
As a measure of this difference,
we determine the ratio, {$\rch$},
of the mean gluon to quark
jet charged particle multiplicity for
{\mysph}$\,\leq\,$1.
Our measurement of this ratio is
\begin{eqnarray}
  \label{eq-rch}
  \rchysphone & = &  1.919\pm 0.047\,(\mathrm{stat.})
        \pm 0.095\,(\mathrm{syst.})  \;\;\;\; .
\end{eqnarray}
For this ratio,
the quark jet result has been corrected for the
small difference in energy between the gluon and
quark jets in the manner described above.
The corresponding result for
\mbox{{\mysph}$\,\leq\,$2} is 
$\rchysphtwo$$\,$=$\,$$1.852\pm 0.034\,(\mathrm{stat.})
\pm 0.077\,(\mathrm{syst.})$.
For purposes of comparison,
we also report our measurement of $\rch$
in full phase space.
This result,
$\rch$$\,$=$\,$$1.514\pm 0.019\,(\mathrm{stat.})
\pm 0.034\,(\mathrm{syst.})$,
agrees well with our previous 
measurements~\cite{bib-opalqg96,bib-opalqg97}
and with recent QCD calculations
of this quantity~\cite{bib-lupia}.
These results are summarized in Table~\ref{tab-rch}.

For completeness,
we also update our measurement of the mean charged
particle multiplicity in {\gincl} jets.
We obtain
{\mngincl}$\,$=$\,$$14.28\pm 0.18\,(\mathrm{stat.})\pm 0.31\,(\mathrm{syst.})$,
in agreement with our earlier results~\cite{bib-opalqg96,bib-opalqg97}
but with a reduced uncertainty.
The corresponding results for \mbox{{\mysph}$\,\leq\,$2}
and \mbox{{\mysph}$\,\leq\,$1}
are included in the bottom rows of
Tables~\ref{tab-nchylttwo} and~\ref{tab-nchyltone}.

The emphasis in the current study is on the
multiplicity of soft particles in jets.
In contrast,
we previously studied the {\it total} charged multiplicity
in gluon and quark jets~\cite{bib-opalqg96,bib-opalqg97}.
For {\it soft} particles,
i.e.~particles with energies $E$$\,<<\,${\ejet},
QCD predicts that the mean multiplicities
in gluon and quark jets
differ by a factor of 
$r$$\,$=$\,${\cacf}$\,$=$\,$2.25~\cite{bib-stan,bib-stanprivate}.
Because our experimental definition of jets
corresponds to the theoretical one,
our result (relation~(\ref{eq-rch}))
provides the most direct test of this prediction to date.
Nonetheless, 
the QCD prediction refers to partons whereas
the measurement is based on hadrons.
Furthermore, the QCD result
does not account for energy-momentum conservation
or higher order perturbative terms.
These latter corrections are believed to be negligible
in the asymptotic limit $E$$\,<<\,${\ejet}~\cite{bib-stanprivate}.
Thus the directness of our test is limited only
by hadronization effects and the extent to which
the asymptotic condition is satisfied by our data.

To demonstrate the
correspondence between our experimental
variable~(\ref{eq-rch}) and $r$ as it is defined
for analytic calculations,
and to assess the origin of the
remaining difference between
our measurement ({$\rch$$\,\approx\,$1.92}
for $\mmysph$$\,\leq\,$1)
and the QCD prediction ($r$$\,$=$\,$2.25),
we examined the predictions of the Herwig Monte Carlo
at the hadron and parton levels and for
{\ejet}$\,$=$\,$40.1~GeV (as in our analysis) and {\ejet}$\,$=$\,$5~TeV.
The parton level results are obtained using the final-state partons,
i.e., those which are present after termination of the parton shower.
Herwig incorporates exact energy-momentum conservation,
higher order perturbative terms up to and beyond the
next-to-next-to-leading order,
a hadronization model,
and exhibits the correct QCD asymptotic behavior
as the c.m. energy becomes large
(see, for example,
Fig.~2 in~\cite{bib-opalqg96}).\footnote{In contrast,
Jetset does {\it not} exhibit the correct 
asymptotic behavior~\cite{bib-opalqg96}.}
Herwig is thus well suited to compare both to our data
and to the asymptotic QCD result.
Specifically,
we determine the Herwig prediction for
$\rch$({\mysph}$\,\leq\,$1) and the corresponding result
$r$({\mysph}$\,\leq\,$1) at the parton level.
To determine these ratios,
we use event hemispheres 
in Herwig {\gluglu} and uds events.
At the hadron level,
Herwig predicts {$\rch$({\mysph}$\,\leq\,$1)}
to be 1.92 for {\ejet}$\,$=$\,$40.1~GeV
and 2.18 for {\ejet}$\,$=$\,$5~TeV
(the statistical uncertainties of the Monte Carlo
results are negligible).
At the parton level,
the corresponding results for
$r$({\mysph}$\,\leq\,$1) are 2.06 and~2.25.
The hadron level result for 40.1~GeV 
is in good agreement with our measurement
(relation~(\ref{eq-rch})).
The parton level result for 5~TeV jets yields precisely
the QCD asymptotic value of~2.25 demonstrating
that our experimental variable $\rch$({\mysph}$\,\leq\,$1)
does indeed correspond to~$r$ as it is defined analytically.
Thus,
our data are consistent with the QCD prediction.
The hadronization corrections,
given by the ratios of the parton to the hadron level
Monte Carlo results,
are 1.07 for {\ejet}$\,$=$\,$40.1~GeV and 1.03
for {\ejet}$\,$=$\,$5~TeV.
The corrections for finite energy,
given by the ratios of the Herwig results at 5~TeV
to those at 40.1~GeV,
are 1.14 at the hadron level and 
1.09 at the parton level.
We conclude that the difference between our measurement
($\rch$({\mysph}$\,\leq\,$1){$\,\approx\,$1.92}) 
and the QCD prediction ($r$$\,$=$\,$2.25)
can mostly be explained by the effects of finite energy.

\subsection{Mean multiplicity ratio of 
soft particles at large \boldmath{$\pperp$}}
\label{sec-softpt}

It was recently
noted~\cite{bib-ochs} that the multiplicity ratio of
gluon to quark jets should exhibit a value near
the full asymptotic prediction of {\cacf}$\,$=$\,$2.25,
even at the finite energies of LEP,
if soft particles with large transverse
momenta to the jet axes are considered:
soft gluons at large angles to the jet axes
in gg or {\qq} events are emitted coherently,
with a coupling strength proportional to the
effective color charge of the parton initiating the jet,
given by {\cca} for gluon jets
and {\ccf} for quark jets.
We therefore examined the predictions of the Herwig
Monte Carlo for the ratio of the $\pperp$ distributions
of gluon to quark jets for soft charged hadrons defined by
$p$$\,<\,$4.0~GeV/$c$.
We used hemispheres in Herwig gg and uds events to define the jets,
with {\ejet}$\,$=$\,$40.1~GeV to correspond to our data.
We chose Herwig for this study for the reasons outlined
in Sect.~\ref{sec-nchrap}.
The results are shown by the solid curve
in Fig.~\ref{fig-ptratios}.
For values of $\pperp$ below about 0.2~GeV/$c$,
the ratio of the gluon to quark jet multiplicity is predicted
to have a value near~1.5.
This ratio increases to approximately 2.25 for
$\pperp$$\,\approx\,$1~GeV/$c$ and remains near
this value for larger~$\pperp$.
Analogous results are obtained for soft hadrons defined by
$p$$\,<\,$2.0~GeV/$c$ and $p$$\,<\,$1.0~GeV/$c$
(dashed and dotted curves),
i.e.~the curves reach values near 2.25 for $\pperp$ values
above about 1.0~GeV/$c$ irrespective of the precise definition
of soft particles.
If all charged particles are selected,
and not just soft ones,
the predicted multiplicity ratio
reaches a maximum of only about 1.85,
however (dash-dotted curve).
The results of Fig.~\ref{fig-ptratios}
suggest that the multiplicity ratio of soft hadrons
at large~$\pperp$ effectively yields a
measurement of {\cacf} at LEP~\cite{bib-ochs}.
It is for this reason that we include the $\pperp$ spectrum of
charged particles with $p$$\,<\,$4.0~GeV/$c$ in our study
(see Sect.~\ref{sec-distr}).
The experimental data for this distribution, $\ptsoft$,
were previously presented in Fig.~\ref{fig-ptsoft}
and Table~\ref{tab-ptsoft}.

As a measure of the soft particle multiplicity at
large~$\pperp$,
we integrate the $\ptsoft$ distribution
between $\pperp$ values of 0.8 and 3.0~GeV/$c$:
this range is chosen on the basis
of Fig.~\ref{fig-ptratios},
as the region where the ratio of the gluon to quark
jet multiplicity is predicted to
approximately equal~{\cacf}.
We choose the upper limit of integration to be 
$\pperp$$\,$=$\,$3.0~GeV/$c$ to avoid the region near the
kinematic boundary at $\pperp$$\,$=$\,$4~GeV/$c$
(in practice this makes little difference because of the small
statistical weight of particles with $\pperp$$\,>\,$3~GeV/$c$).
Our measurement of this quantity, $\rchsoft$, is:
\begin{eqnarray}
  \label{eq-rchsoft}
  \rchsoft & = &  2.29\pm 0.09\,(\mathrm{stat.})
        \pm 0.15\,(\mathrm{syst.})  \;\;\;\; .
\end{eqnarray}
This result is summarized in Table~\ref{tab-rch}.
The corresponding results from Herwig
at the hadron and parton levels
are 2.16 and 2.09 with
negligible statistical uncertainties.
The hadron level result agrees with our 
measurement~(\ref{eq-rchsoft}) to within the 
experimental uncertainties.
For 5~TeV jets,
Herwig predicts 2.23 and 2.25 at the hadron 
and parton levels:
the latter result equals
the QCD asymptotic prediction for~$r$,
demonstrating the correspondence between 
the variable~(\ref{eq-rchsoft}) and $r$ as it
is defined analytically,
similar to the variable $\rch$({\mysph}$\,\leq\,$1)
considered in Sect.~\ref{sec-nchrap}.
The ratio~(\ref{eq-rchsoft}) has smaller 
predicted corrections than the variable
considered in Sect.~\ref{sec-nchrap}:
the hadronization correction for
$\rchsoft$ at 40.1~GeV is 0.97
(compared to 1.07 for $\rch$({\mysph}$\,\leq\,$1)),
while the correction for finite energy
between 40.1~GeV and 5~TeV hadron jets
is 1.03 (compared to 1.14).
Thus the ratio of
soft hadron multiplicities between gluon and quark jets
at large~$\pperp$ does indeed
yield a value consistent with {\cacf}
to within the uncertainties,
even at the finite energies of LEP,
as predicted in~\cite{bib-ochs}.

For completeness,
we also report the results we obtain 
for the gluon to quark jet multiplicity ratio
using different choices for the intervals of $p$ and~$\pperp$.
For $p$$\,<\,$2~GeV/$c$ (rather than $p$$\,<\,$4~GeV/$c$)
we obtain
$\rchsofttwogev$$\,$=$\,$$2.32\pm 0.12\,(\mathrm{stat.})
\pm 0.14\,(\mathrm{syst.})$.
For 0.8$\,<\,$$\pperp$$\,<\,$2.0~GeV/$c$
(rather than 0.8$\,<\,$$\pperp$$\,<\,$3.0~GeV/$c$)
we obtain
$\rchsoftpttwo$$\,$=$\,$
$2.33\pm 0.08\,(\mathrm{stat.})
\pm 0.16\,(\mathrm{syst.})$.
These results are very similar to
the result~(\ref{eq-rchsoft}) reported above.

\subsection{Fragmentation function}

Another striking feature of our results is the much
softer fragmentation function of gluon jets compared
to quark jets (Fig.~\ref{fig-xee}).
That gluon jets have a softer fragmentation function
than quark jets has already been well established.
These earlier studies either employed a jet finder to define
the gluon jets~\cite{bib-qg95a,bib-qg93,bib-qgff}
or extracted the gluon jet fragmentation function using
measurements of the longitudinal and transverse
fragmentation functions in {\epem} 
annihilations~\cite{bib-opaltransverse,bib-transverse}.
Unlike the earlier studies based on jet finders,
we employ theoretically well defined jets.
Unlike the earlier studies utilizing the longitudinal 
and transverse fragmentation functions,
the energy scale of our jets is well defined.
Therefore our results have more theoretical meaning
than these previous ones.

\section{Monte Carlo predictions}
\label{sec-mc}

Figs.~\ref{fig-rapidity}--\ref{fig-nchy} include
the hadron level predictions of Jetset, Herwig and Ariadne.
The Monte Carlo results for {\gincl} jets are
obtained using parton level information
in the manner described in Sect.~\ref{sec-gluon}.
The results for Ariadne are shown both with and without
the effects of color reconnection.
For the standard version of Ariadne,
i.e.~the version {\it without} reconnection,
we use the parameter values given in~\cite{bib-alephes}
with the following modification:
the value of the ``{\it a}'' parameter\footnote{Given
by the Jetset Monte Carlo parameter PARJ(41).}
controlling the hardness of the fragmentation function
is increased from 0.40 to 0.52 to obtain a better 
description of {\mnch} in inclusive
multihadronic Z$^0$ events.
These parameters provide a substantial improvement
in the description of our uds jet data
compared to the default parameters.
We examine two versions of the Ariadne model {\it with} 
reconnection,
referred to here as {\artwo} and {\arthree}
to conform to our previous usage~\cite{bib-nige.}.
In the {\artwo} model\footnote{Enabled by
setting the parameter MSTA(35)$\,$=$\,$2.},
gluons are not subject to
reconnection (see Sect.~\ref{sec-reconnection})
unless their energies are below a 
cutoff.\footnote{Given by the parameter PARA(28).}
For our study,
this cutoff is set to~2~GeV~\cite{bib-lonnblad}.
We generate events using the parameters in~\cite{bib-alephes}
except for the {\it a} parameter which is adjusted to 0.65
to obtain an accurate description of {\mnch} 
in inclusive Z$^0$ events.
In the {\arthree} model\footnote{Enabled by 
setting the parameter MSTA(35)$\,$=$\,$3.},
gluons of all energies are subject to reconnection.
For this model we use the parameters in~\cite{bib-alephes}
except with the {\it a} parameter set to 0.58
to describe {\mnch} in inclusive Z$^0$ events.
The results for the
mean charged particle multiplicity in inclusive hadronic Z$^0$ events
are 20.9, 20.9 and 21.0 for our tuned versions of {\artwo}, {\arthree}
and the standard version of Ariadne, respectively,
in agreement with the measured value of 
$21.0\pm0.2$~\cite{bib-opaltransverse}-\cite{bib-alephes}.

The three versions of Ariadne yield very similar
descriptions of standard measures of 
event properties in inclusive Z$^0$ multihadronic events,
such as thrust, sphericity, aplanarity
(see e.g.~\cite{bib-opal90} for a definition of these variables)
and the quantities defined in Sect.~\ref{sec-distr}.
Their overall descriptions of inclusive Z$^0$ data are good.
To illustrate these points,
we calculated the $\chi^2$ values between the predictions
of the models and the measured distributions 
of thrust T~\cite{bib-opthrust}, jet broadening
variable B$_{\mathrm{W}}$~\cite{bib-opthrust},
scaled particle momentum 
$x_p$$\,$=$\,$$2p$/{\ecm}~\cite{bib-opalhapke},
rapidity with respect to the sphericity axis~\cite{bib-delys},
and charged particle multiplicity in the rapidity
interval \mbox{$\mmysph\leq 2$}~\cite{bib-alephnch},
for inclusive hadronic Z$^0$ events.
(Note that there are correlations between these
variables and between different bins of some of
the distributions.)
These last three variables are chosen because of their
similarity to distributions studied in this paper
(Sect.~\ref{sec-distr}).
The total $\chi^2$ values for 152 bins of data are
293 for {\artwo}, 241 for {\arthree},
and 290 for the standard version of Ariadne.
For purposes of comparison,
the corresponding result from Jetset is~322.
In Figs.~\ref{fig-thrust_bw_xp_z0incl} 
and~\ref{fig-rapidity_nchylt2_z0incl},
the predictions of the three versions of Ariadne
are shown in comparison to the inclusive
Z$^0$ event measurements.
The predictions of the three variants of Ariadne 
are seen to be virtually
indistinguishable from each other and in good
agreement with the data.

From Figs.~\ref{fig-rapidity}--\ref{fig-nchy},
it is seen that the Monte Carlo simulations provide a
good description of the gluon jet properties,
with the exception 
of the {\artwo} and {\arthree} color reconnection models
whose predictions for {\gincl} jets
are discussed in the next section.
All the models provide a reasonable description of
the uds jet measurements.
The predictions of the {\artwo} and {\arthree} models for
the uds jet properties are essentially identical to those
of the standard version of Ariadne.

The Monte Carlo predictions for the multiplicity
ratios $\rch$, $\rchysphtwo$, $\rchysphone$ and $\rchsoft$
are given in Table~\ref{tab-rch}.
In addition to the results for charged hadrons,
the results are given at the parton level (in parentheses)
for Herwig, Jetset and the standard version of Ariadne.
The hadron level predictions are in general agreement
with the data,
with the exception of the Jetset prediction for
$\rchsoft$ which is about 2 standard deviations of
the total experimental uncertainty below the measurement
(this is possibly related to the failure of Jetset to yield
the QCD asymptotic result r$\,$=$\,${\cacf} at large
jet energies~\cite{bib-opalqg96}).
Also included in Table~\ref{tab-rch}
are the predictions of a special version of
Jetset in which the color factor {\cca} governing gluon jet
evolution has been set equal to the factor {\ccf}$\,$=$\,$4/3 governing
quark jet evolution.
The parton level results with {\cca}$\,$=$\,${\ccf}$\,$=$\,$4/3
essentially equal unity,
i.e.~the ratio values vary from 1.00 to 1.06,
in contrast to the parton level predictions of the standard version
of Jetset which range from 1.35 to 1.62,
emphasizing the sensitivity of the multiplicity ratios
to the value of {\cacf}.
At the hadron level,
the multiplicity ratios obtained using
this special version of Jetset
vary between 1.31 and~1.54:
the results at the hadron level are not
expected to equal unity,
even with {\cca}$\,$=$\,${\ccf},
because the Jetset hadronization model
treats quarks and gluons differently.\footnote{Quarks are
attached to a single string segment in the Lund model
of hadronization~\cite{bib-lund} implemented in Jetset,
whereas gluons are attached to two string segments.}

\section{Test of a model for color reconnection}
\label{sec-reconnection}

Most implementations of QCD,
including those in the standard versions of
Jetset, Herwig and Ariadne,
are based on the so-called large 
N$_{\mathrm{c}}$ approximation,
with N$_{\mathrm{c}}$ the number of colors.
In this approximation,
the manner in which partons are connected 
to form an overall color singlet is uniquely specified.
For example,
in Z$^0\rightarrow\,${\qq}gg events,
in which two gluons gg are radiated from a quark q
and antiquark $\overline{\mathrm{q}}$
produced from a Z$^0$ decay,
the quark is color-connected 
to one of the gluons
(e.g.~connected by a color flux tube,
which is modelled as a cluster chain or string in
the Monte Carlo programs),
this first gluon is color-connected to the second gluon,
and the second gluon is color-connected to
the antiquark~$\overline{\mathrm{q}}$.
Thus, the entire event consists of
a single color singlet system.
This color singlet hadronizes,
with hadrons appearing preferentially in the regions
spanned by the color flux.

In the large N$_{\mathrm{c}}$ approximation,
interference terms of relative order 
1/N$_{\mathrm{c}}^2$ are ignored.
If these interference terms are included
to obtain predictions valid
beyond the large N$_{\mathrm{c}}$ approximation,
the manner in which partons are connected
to each other is no longer specified uniquely.
For example,
in Z$^0\rightarrow\,${\qq}gg events,
the possibility 
that the q and $\overline{\mathrm{q}}$
form a color singlet by themselves,
with the two gluons gg forming a separate color singlet,
occurs with probability\footnote{In addition, 
dynamical effects can lead to a
further suppression of these ``reconnected'' terms.}
1/(N$_{\mathrm{c}}^2$-1)
relative to the ``normal'' situation described
in the previous paragraph.
The possibility of defining the color singlets in this
latter manner is an example of what is
called color reconnection.
Color reconnection can affect events at both
the perturbative and non-perturbative levels:
its effects at the perturbative level are expected
to be small, however,
in comparison to its effects at the non-perturbative
level~\cite{bib-ks1}.
To assess the effects of color reconnection
at the non-perturbative level,
several models for reconnection have been implemented in 
non-standard versions of QCD Monte Carlo
event generators.
With the exception of the VNI model~\cite{bib-vni} 
discussed in~\cite{bib-nige.},
none of these models has been subjected
to a stringent test.
Color reconnection has been a topic of recent interest
due to the possibility that reconnected diagrams
could measurably affect
the reconstructed W boson mass in
{\epem}$\rightarrow$
W$^+$W$^-\rightarrow\,$
q$_1$$\overline{\mathrm{q}}_2$q$_3$$\overline{\mathrm{q}}_4$
events recorded at LEP-2~\cite{bib-lepww}.

In general,
color reconnection can have a significant influence
on the energy and 
angular distributions of hadrons in an event
since the color flux spans different regions of phase space
compared to normal color connection.
Since the standard Monte Carlo programs provide a good
description of the general properties
of inclusive Z$^0$ hadronic events,
it can be inferred that the overall effect of
reconnection is small.
It is nonetheless possible that the effects of reconnection
are sizable in special classes of events such as the
{\epem}$\rightarrow\,${\qq}$\,${\gincl} events studied here.
Indeed,
it has been suggested~\cite{bib-lepww,bib-gustafson} 
that this class of events
--~with a pure system of gluons recoiling against a
quark-antiquark system in the opposite hemisphere~--
can provide a sensitive test for the presence of
reconnection phenomena.
In the following,
we use our data to test the Ariadne 
color reconnection models {\artwo} and \mbox{{\arthree}}
presented in Sect.~\ref{sec-mc}.
We choose these models for our study because they
provide good descriptions of inclusive Z$^0$ data,
as discussed in Sect.~\ref{sec-mc},
and thus represent realistic models of nature
(unlike the VNI model of reconnection
which does not describe the basic properties of
W$^+$W$^-$ events~\cite{bib-nige.}).

\begin{figure}[ht]
\epsfxsize=17cm
\epsffile[65   100 595 700]{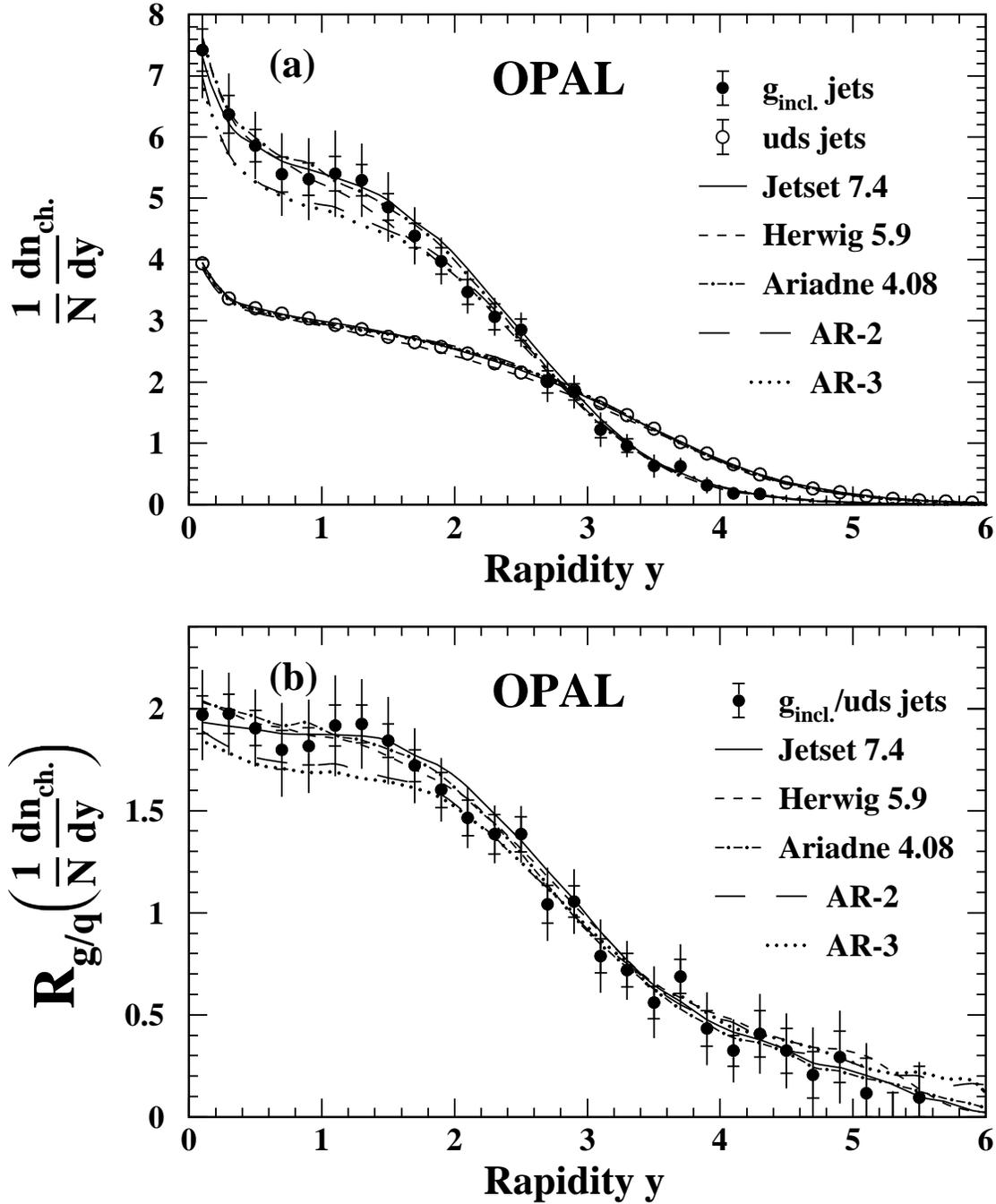}
\caption{
(a)~Corrected distributions of charged particle rapidity, {\ysph},
for 40.1~GeV {\gincl} gluon jets
and 45.6~GeV uds quark jets.
(b)~The ratio of the gluon to quark jet rapidity
distributions for 40.1~GeV jets.
The total uncertainties are shown by vertical lines.
The experimental statistical uncertainties are
indicated by small horizontal bars.
(The uncertainties are too small 
to be seen for the uds jets.)
The predictions of various parton shower 
Monte Carlo event generators are also shown.
These data are tabulated in Table~\ref{tab-rapidity}.
}
   \label{fig-rapidity}
\end{figure}

\begin{figure}[ht]
\epsfxsize=17cm
\epsffile[65   100 595 700]{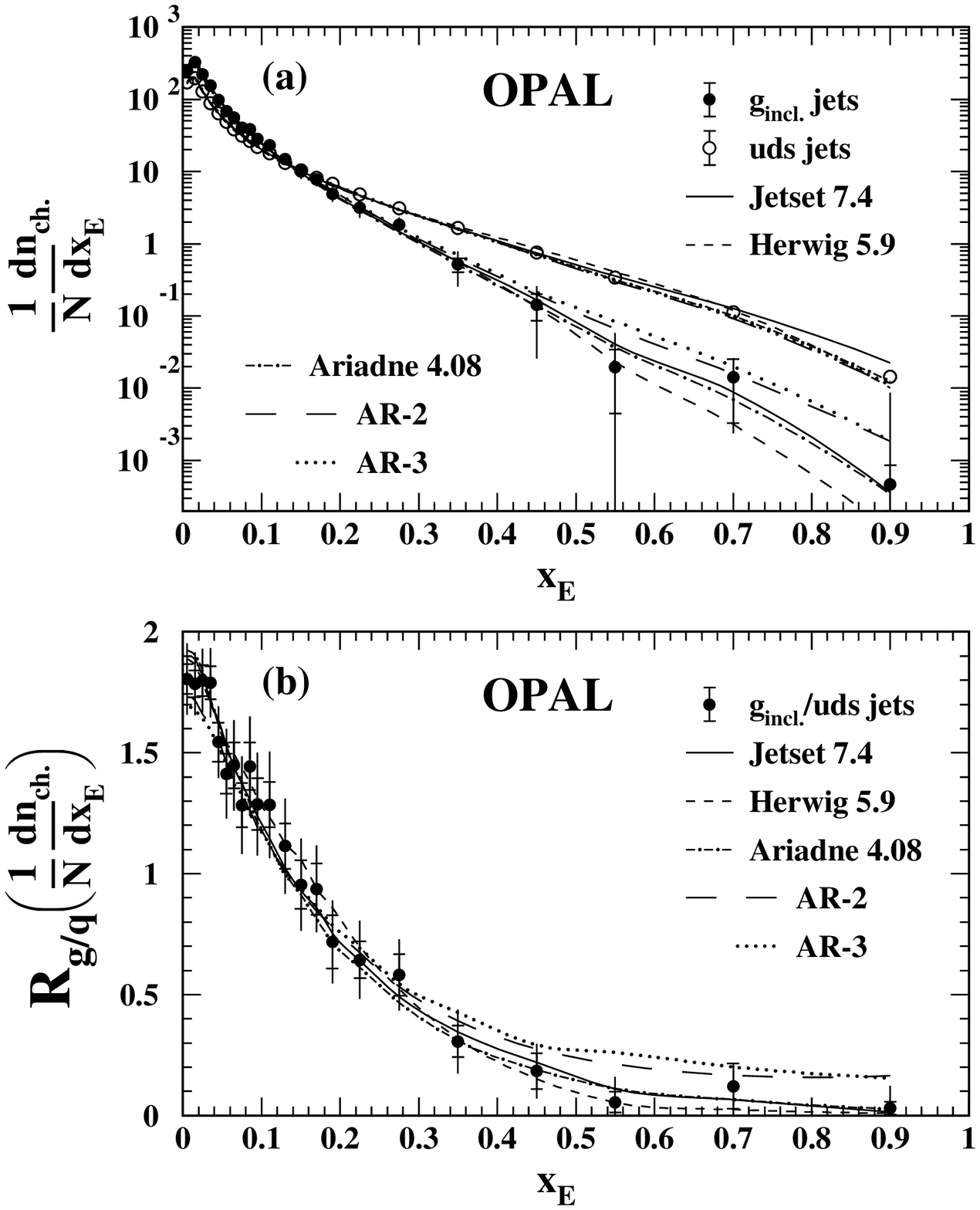}
\caption{
(a)~Corrected distributions of charged particle
scaled energy, {\xee}$\,$=$\,$$E/E_{\,\mathrm{jet}}$,
for 40.1~GeV {\gincl} gluon jets
and 45.6~GeV uds quark jets.
(b)~The ratio of the gluon to quark jet {\xee}
distributions for 40.1~GeV jets.
The total uncertainties are shown by vertical lines.
The experimental statistical uncertainties are
indicated by small horizontal bars.
(The uncertainties are too small 
to be seen for the uds jets.)
The predictions of various parton shower 
Monte Carlo event generators are also shown.
These data are tabulated in Table~\ref{tab-xee}.
}
   \label{fig-xee}
\end{figure}

\begin{figure}[ht]
\epsfxsize=17cm
\epsffile[65   100 595 700]{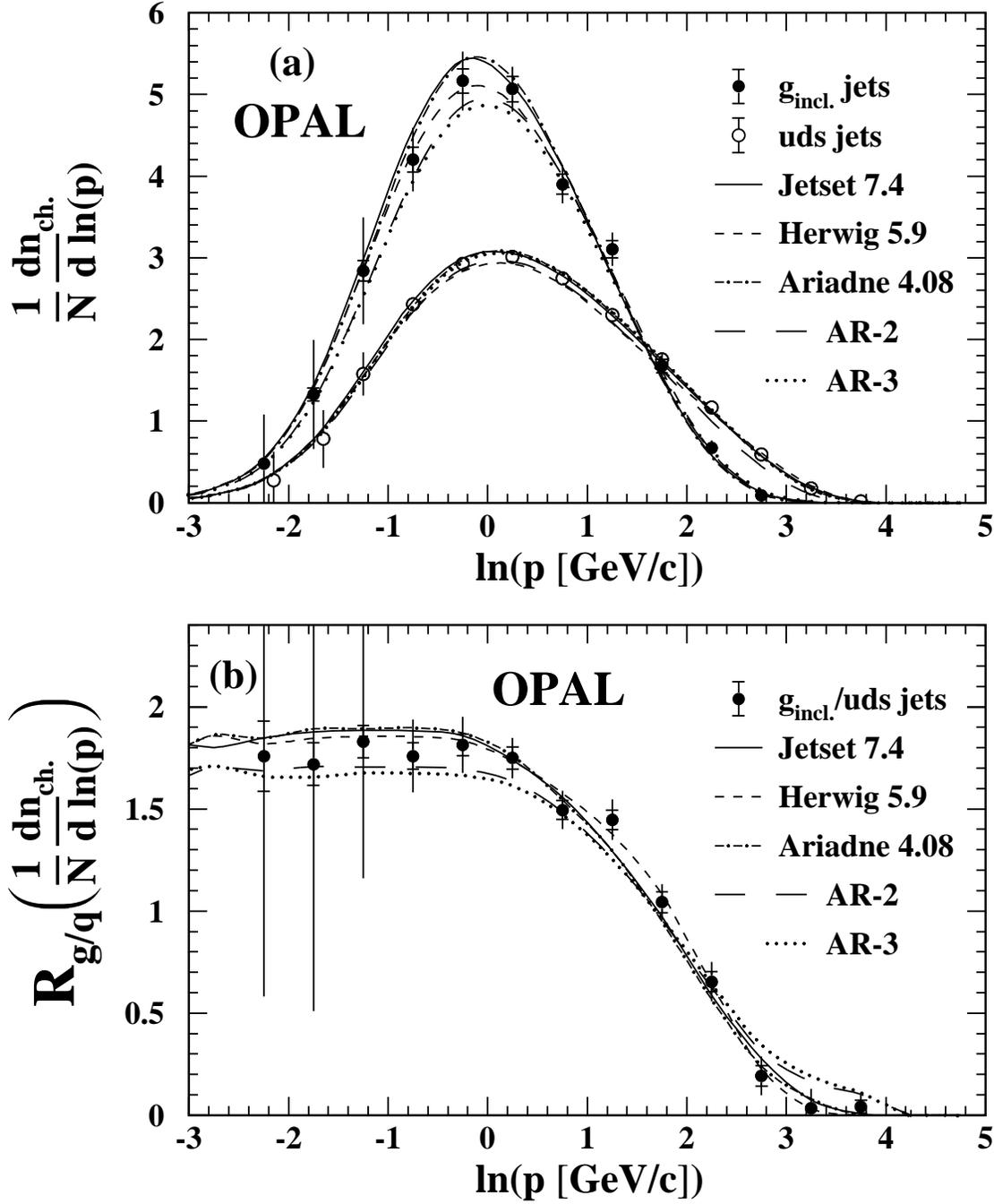}
\caption{
(a)~Corrected distributions of 
the logarithm of charged particle momentum,
$\ln\,(p)$,
for 40.1~GeV {\gincl} gluon jets
and 45.6~GeV uds quark jets.
(b)~The ratio of the gluon to quark jet $\ln\,(p)$
distributions for 40.1~GeV jets.
The total uncertainties are shown by vertical lines.
The experimental statistical uncertainties are
indicated by small horizontal bars.
(The statistical uncertainties are too small 
to be seen for the uds jets.)
The predictions of various parton shower 
Monte Carlo event generators are also shown.
These data are tabulated in Table~\ref{tab-lnp}.
}
   \label{fig-lnp}
\end{figure}

\begin{figure}[ht]
\epsfxsize=17cm
\epsffile[65   100 595 700]{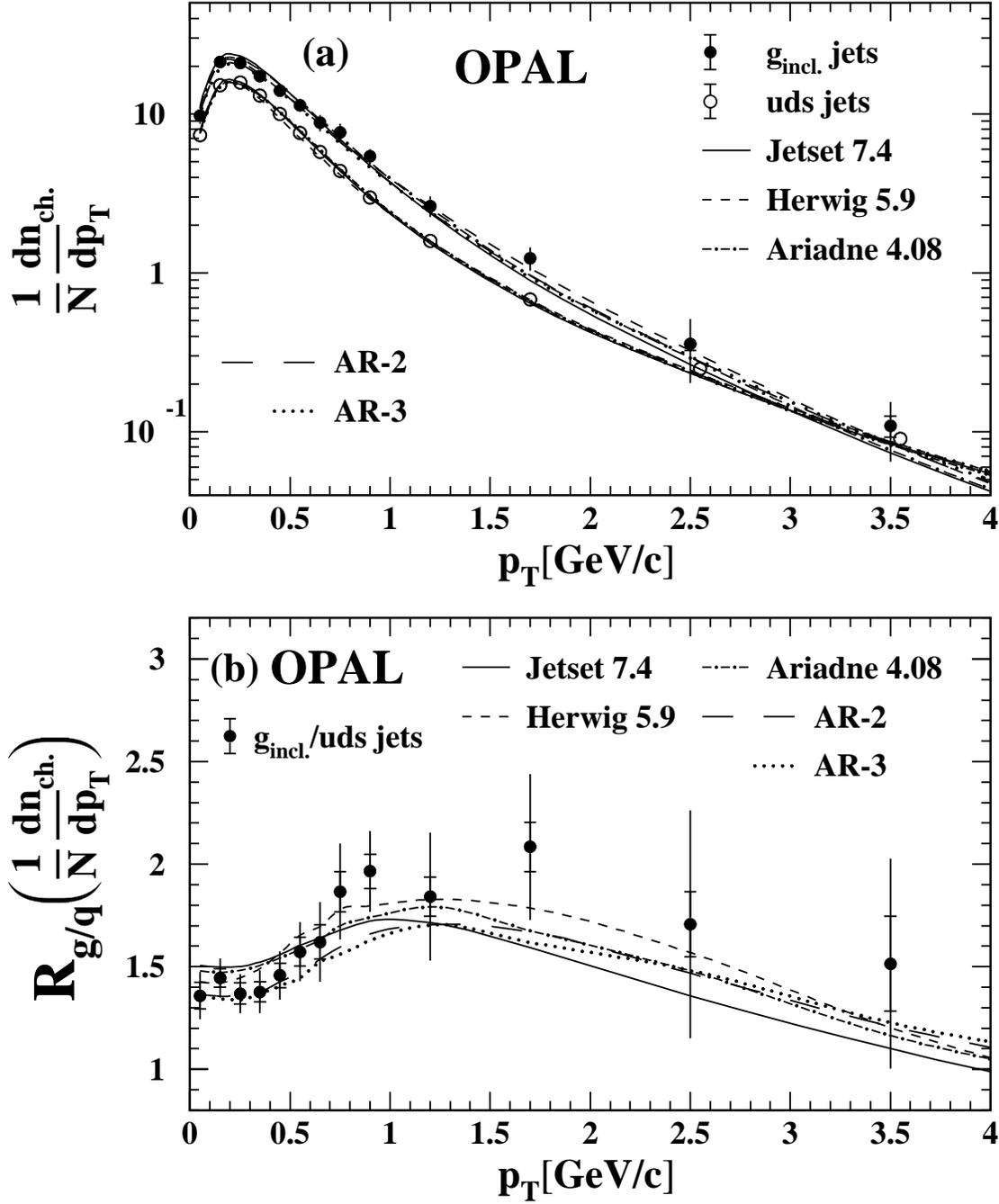}
\caption{
(a)~Corrected distributions of charged particle
transverse momentum with respect to the jet axis,
$\pperp$,
for 40.1~GeV {\gincl} gluon jets
and 45.6~GeV uds quark jets.
(b)~The ratio of the gluon to quark jet $\pperp$
distributions for 40.1~GeV jets.
The total uncertainties are shown by vertical lines.
The experimental statistical uncertainties are
indicated by small horizontal bars.
(The uncertainties are too small 
to be seen for the uds jets.)
The predictions of various parton shower 
Monte Carlo event generators are also shown.
These data are tabulated in Table~\ref{tab-ptall}.
}
   \label{fig-ptall}
\end{figure}

\begin{figure}[ht]
\epsfxsize=17cm
\epsffile[65   100 595 700]{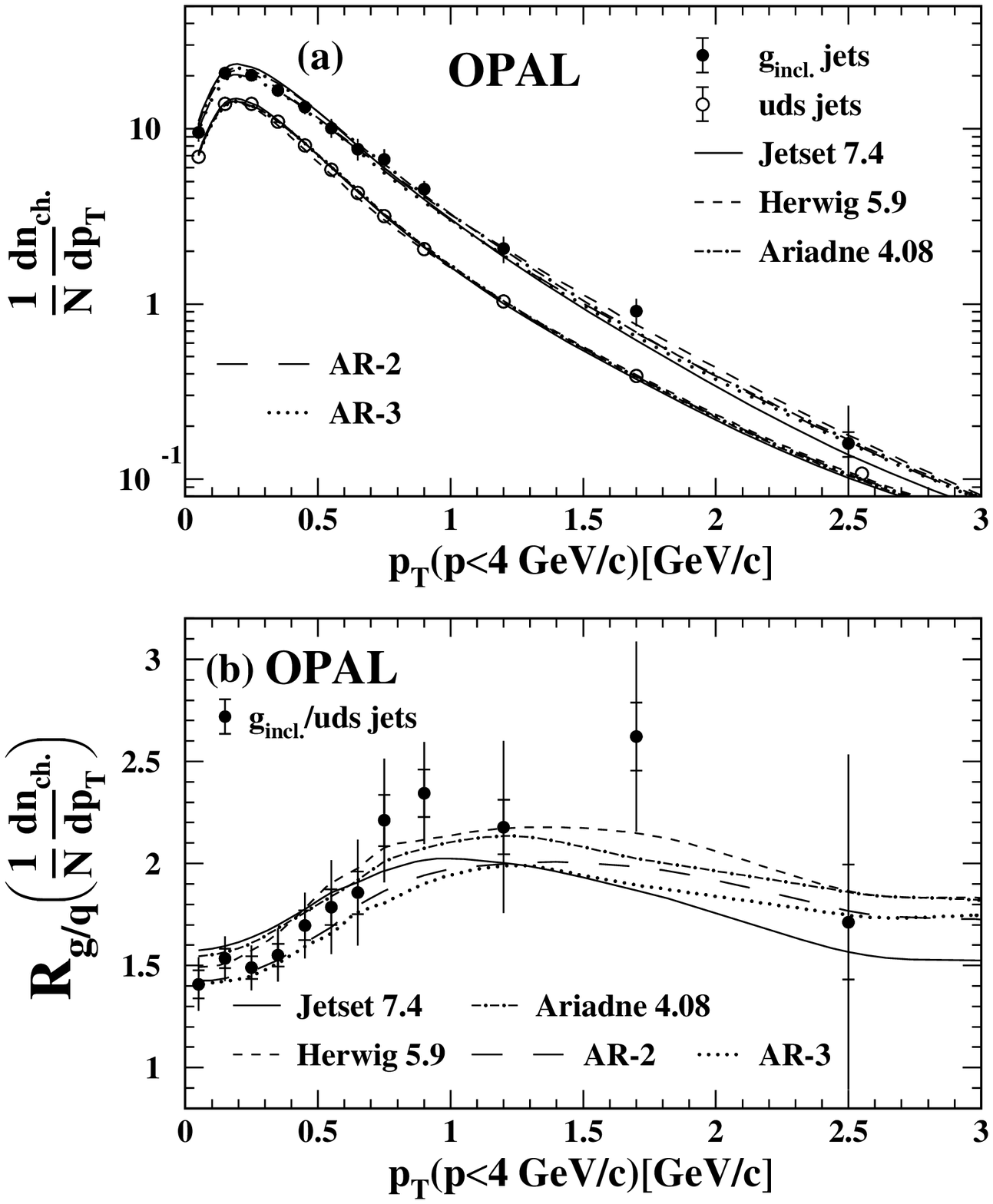}
\caption{
(a)~Corrected distributions of charged particle
transverse momentum with respect to the jet axis
for particles with momentum $p$$\,<\,$4~GeV/$c$,
$\ptsoft$,
for 40.1~GeV {\gincl} gluon jets
and 45.6~GeV uds quark jets.
(b)~The ratio of the gluon to quark jet $\ptsoft$
distributions for 40.1~GeV jets.
The total uncertainties are shown by vertical lines.
The experimental statistical uncertainties are
indicated by small horizontal bars.
(The uncertainties are too small 
to be seen for the uds jets.)
The predictions of various parton shower 
Monte Carlo event generators are also shown.
These data are tabulated in Table~\ref{tab-ptsoft}.
}
   \label{fig-ptsoft}
\end{figure}

\begin{figure}[ht]
\epsfxsize=17cm
\epsffile[65   100 595 700]{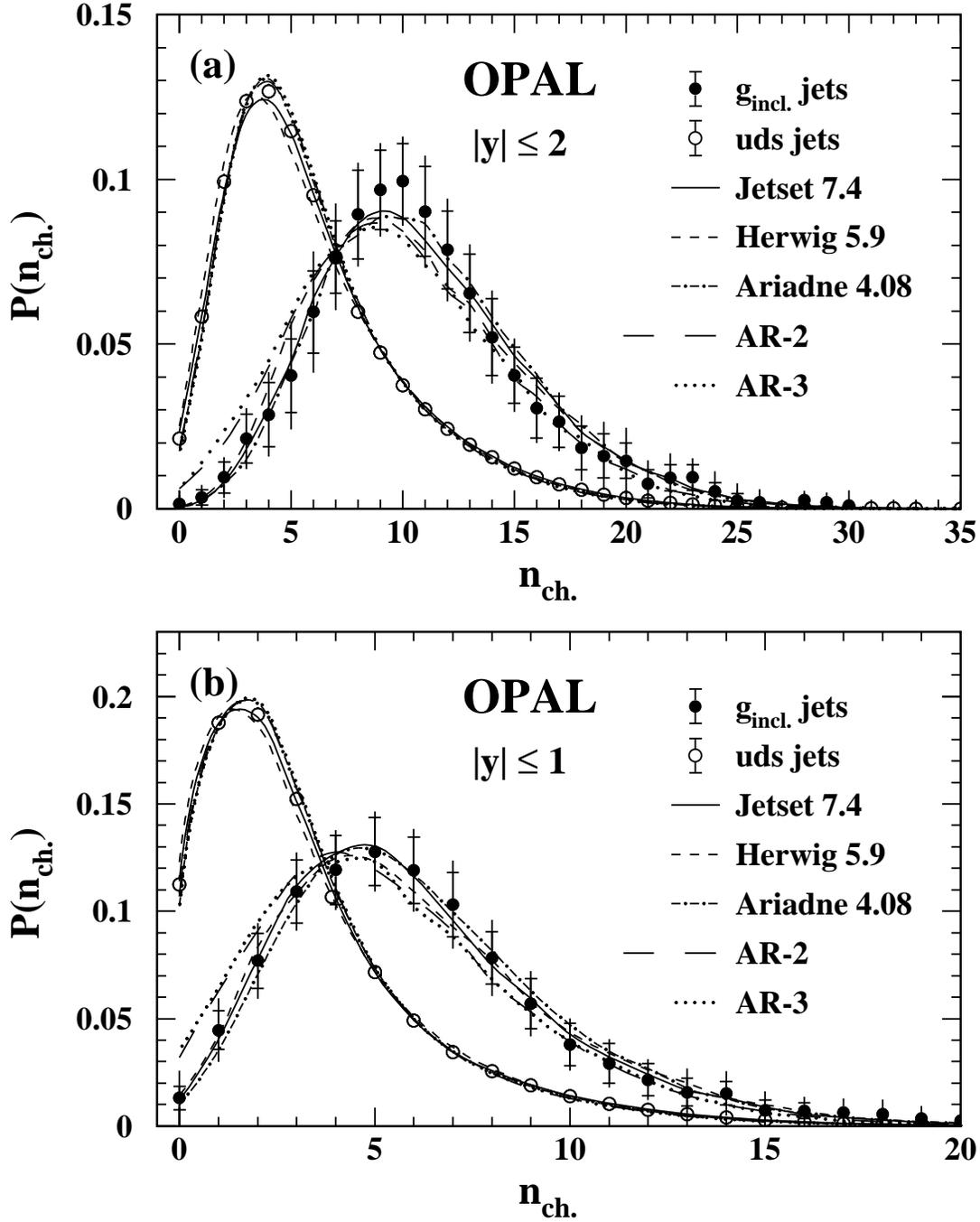}
\caption{Corrected distributions of charged particle
multiplicity in the rapidity intervals
(a)~{\mysph}$\,\leq\,$2 and (b)~\mbox{{\mysph}$\,\leq\,$1}
for 40.1~GeV {\gincl} gluon jets
and 45.6~GeV uds quark jets.
The total uncertainties are shown by vertical lines.
The experimental statistical uncertainties are
indicated by small horizontal bars.
(The statistical uncertainties are too small 
to be seen for the uds jets.)
The data are correlated between bins.
The predictions of various parton shower 
Monte Carlo event generators are also shown.
These data are tabulated in Tables~\ref{tab-nchylttwo}
and~\ref{tab-nchyltone}.
}
   \label{fig-nchy}
\end{figure}

\begin{figure}[ht]
\epsfxsize=17cm
\epsffile[65   100 595 700]{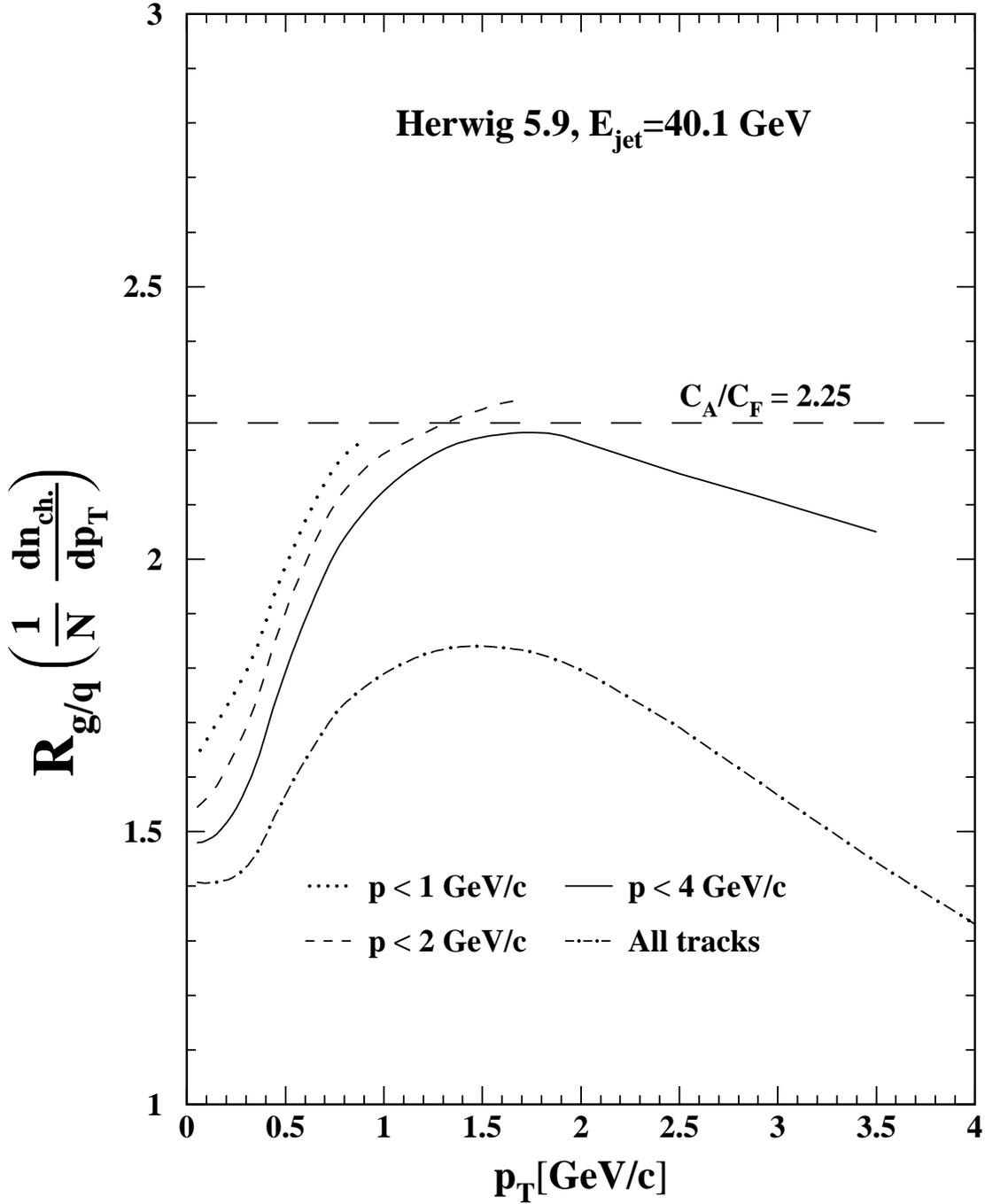}
\caption{
The prediction of the Herwig Monte Carlo event generator
for the ratio of the transverse momentum
distributions between 40.1~GeV gluon and 40.1~GeV uds quark jets,
for all charged particles and for charged particles with
momentum $p$ below 4, 2 and 1~GeV/$c$.
}
   \label{fig-ptratios}
\end{figure}

\begin{figure}[ht]
\epsfxsize=17cm
\epsffile[65   100 595 700]{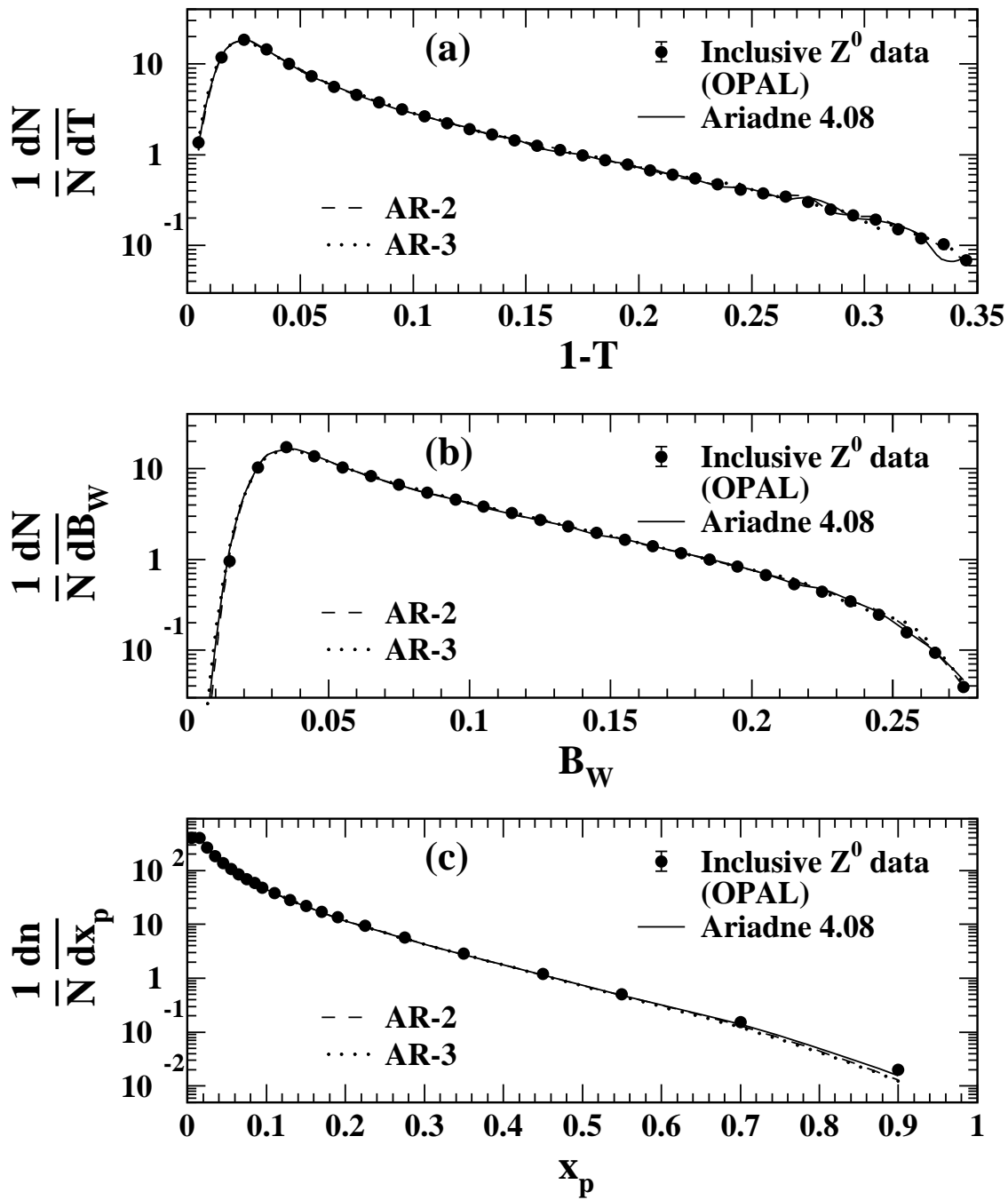}
\caption{
Comparison of the predictions of the standard version of
Ariadne and of the two versions of Ariadne with 
color reconnection to inclusive Z$^0$ event data:
(a)~1-T~[41], with T the thrust,
(b)~jet broadening variable~B$_{\mathrm{W}}$~[41],
and (c)~scaled particle 
momentum~$x_p$$\,$=$\,$$2p$/{\ecm}~[32].
The total uncertainties,
with statistical and systematic terms added in quadrature,
are too small to be visible.
}
   \label{fig-thrust_bw_xp_z0incl}
\end{figure}

\begin{figure}[ht]
\epsfxsize=17cm
\epsffile[65   100 595 700]{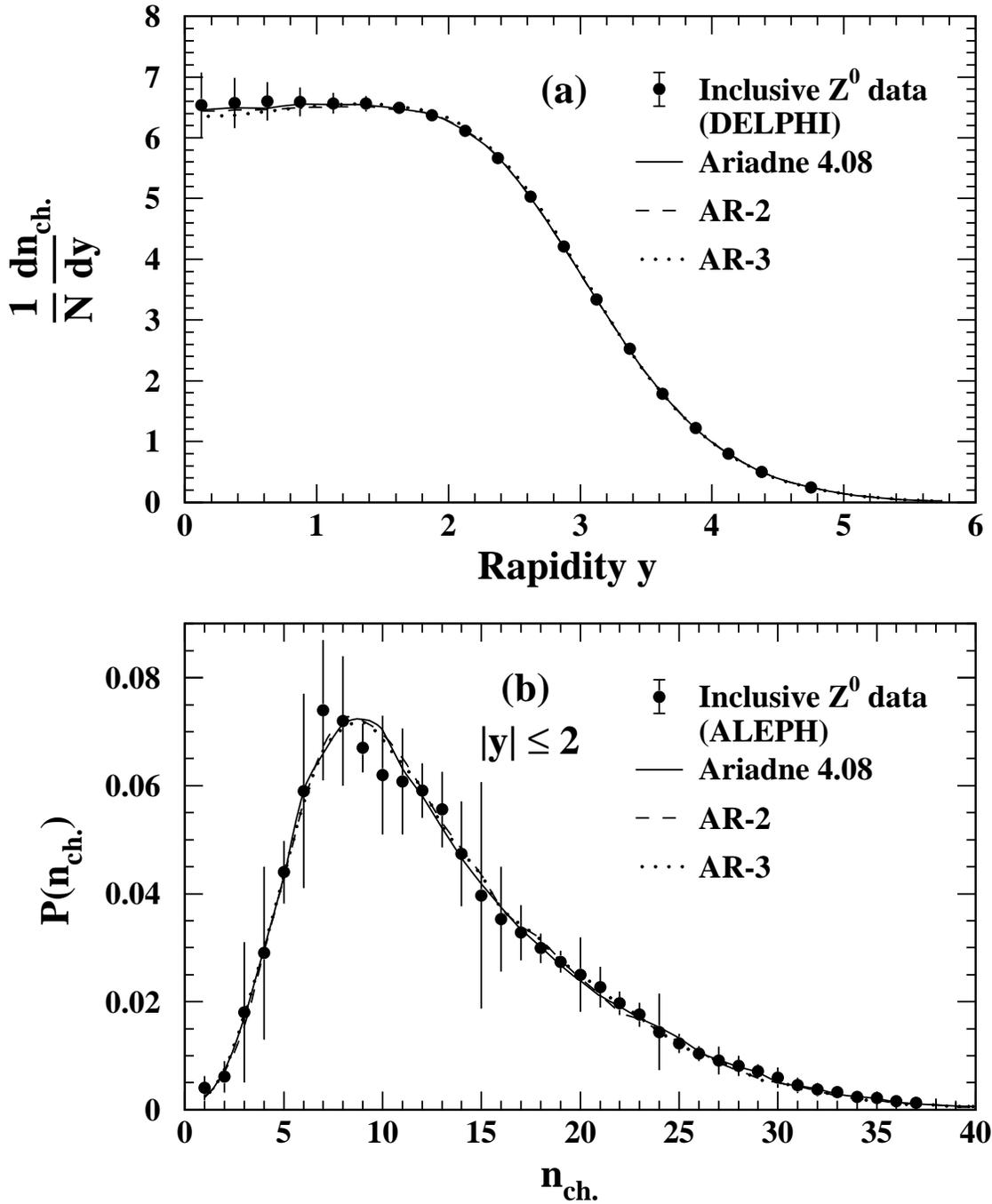}
\caption{
Comparison of the predictions of the standard version of
Ariadne and of the two versions of Ariadne with 
color reconnection to inclusive Z$^0$ event data:
(a)~rapidity {\ysph} with respect to the sphericity axis~[42],
and (b)~charged particle multiplicity in the
interval \mbox{$\mmysph\leq 2$}~[43].
The total uncertainties are shown by the vertical lines,
with statistical and systematic terms added in quadrature.
}
   \label{fig-rapidity_nchylt2_z0incl}
\end{figure}

For {\gincl} jets,
the {\artwo} and {\arthree} models
predict noticeably fewer particles 
at small rapidities and energies,
and noticeably more particles at large
rapidities and energies,
than are observed in either the data or
standard QCD programs
(see Figs.~\ref{fig-rapidity} and~\ref{fig-xee}).
Furthermore,
these two models predict a
downwards shift of about one unit in
the {\gincl} charged particle multiplicity distributions
compared to the data or the
standard QCD programs (see Fig.~\ref{fig-nchy}).
Thus, our {\gincl} data are indeed
sensitive to color reconnection effects.
To test the sensitivity of the models' predictions 
to their parameters,
we varied the values of the main 
parameters\footnote{Specifically,
PARA(1), PARA(3), PARJ(21) and PARJ(42).}
within the uncertainties given in~\cite{bib-alephes}:
the predictions of the models
remained virtually unchanged.
We note that the predictions of the 
{\artwo} and \mbox{{\arthree}}
models are in much more serious
disagreement with our data,
compared to the disagreement seen in
Figs.~\ref{fig-rapidity}, \ref{fig-xee}
and~\ref{fig-nchy},
if the default Ariadne
parameter set is used rather than the parameter
sets described in Sect.~\ref{sec-mc}.

As a quantitative measure of the 
difference between our data
and the predictions of Ariadne with reconnection,
we performed two related tests.
For the first test,
we compared the values of~{$\rch$}.
The results are summarized in the
bottom portion of Table~\ref{tab-rch}:
$\rch$ in full phase space is predicted to be
1.43 and 1.42 by the {\artwo} and {\arthree} models,
which are 2.2 and 2.4 standard deviations of the
total experimental uncertainty below the 
measured value of $1.514\pm0.039\,$(stat.+syst.).
For the second test,
we compared the probability,
measured in per cent,
for a {\gincl} jet to have four or fewer charged
particles with \mbox{$\mmysph\leq 2$}
(a comparison of this nature
is suggested in~\cite{bib-gustafson}).
To determine these probabilities,
we integrated the {\gincl} distributions
in Fig.~\ref{fig-nchy}a from
{\nch}$\,$=$\,$0 to {\nch}$\,$=$\,$4.
The upper limit of {\nch}$\,$=$\,$4 is chosen because it
yields the maximum deviation of the predictions of
{\artwo} and {\arthree} with respect to the 
standard version of Ariadne,
using the statistical uncertainties of the data,
compared to other choices.
The results are given in Table~\ref{tab-nchle5}.
The {\artwo} and {\arthree} models
predict 11.2\% and 12.2\% for these probabilities,
in disagreement with the
measured value of $6.4\pm2.1\,$(stat.+syst.)\%
by 2.3 and 2.8 standard deviations, respectively.
In contrast,
the standard QCD programs reproduce the
experimental result well (Table~\ref{tab-nchle5}).

\begin{figure}[ht]
\epsfxsize=17cm
\epsffile[65   100 595 700]{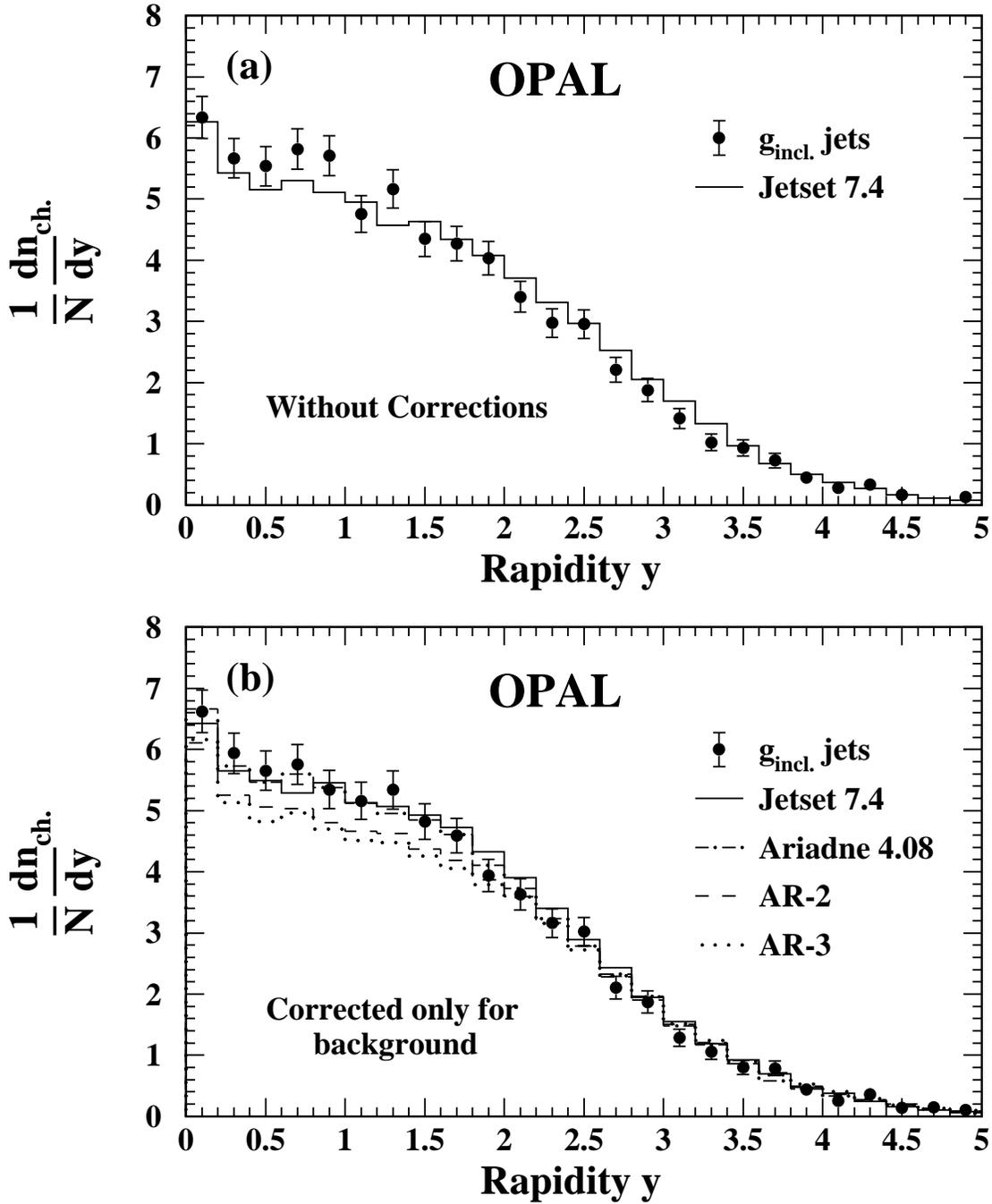}
\caption{
The distribution of charged particle rapidity
for 40.1~GeV {\gincl} gluon jets:
(a)~uncorrected distribution, 
i.e.~at the level which includes background,
detector acceptance and resolution,
secondary interactions,
initial-state radiation,
and the experimental track and cluster
selection criteria,
and (b)~distribution corrected for background only.
The uncertainties are statistical.
}
   \label{fig-uncorry}
\end{figure}

\begin{figure}[ht]
\epsfxsize=17cm
\epsffile[65   100 595 700]{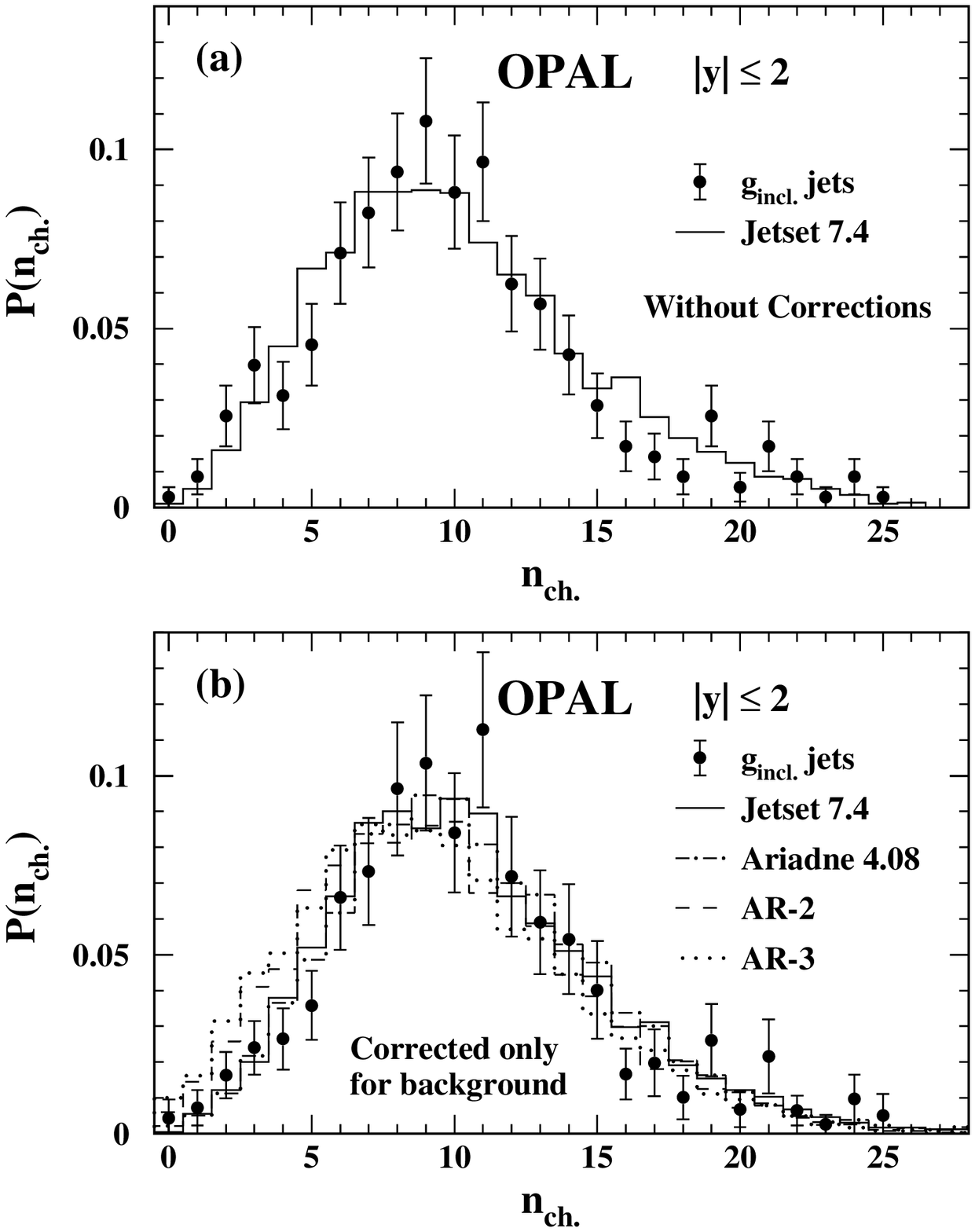}
\caption{
The distribution of charged particle multiplicity in the
rapidity interval \mbox{$\mmysph\leq 2$}
for 40.1~GeV {\gincl} gluon jets:
(a)~uncorrected distribution,
i.e.~at the level which includes background,
detector acceptance and resolution,
secondary interactions,
initial-state radiation,
and the experimental track and cluster
selection criteria,
and (b)~distribution corrected for background only.
The uncertainties are statistical.
}
   \label{fig-uncorrnch}
\end{figure}

The results of the previous paragraph are based
on fully corrected data,
emphasizing the absolute measurement
of gluon jet multiplicity.
By examining the gluon jet properties at 
the level which includes detector
acceptance and resolution and the
experimental selection criteria,
it is possible to emphasize the {\it relative} difference
between data and model
since factors like the experimental track and cluster
definitions are common to both.
Such a comparison is presented in
Figs.~\ref{fig-uncorry} and~\ref{fig-uncorrnch}
for rapidity and 
charged particle multiplicity with~\mbox{$\mmysph\leq 2$}.
In Figs.~\ref{fig-uncorry}a and~\ref{fig-uncorrnch}a,
the predictions of Jetset including detector simulation
and the same analysis procedures as are applied to the
data are shown in comparison to the
experimental measurements without corrections.
Jetset is seen to reproduce the data well,
without significant systematic deviations.
In Figs.~\ref{fig-uncorry}b and~\ref{fig-uncorrnch}b,
the predictions of Jetset and the three versions
of Ariadne are shown after including detector simulation
and the experimental selection criteria of
Sects.~\ref{sec-detector} and~\ref{sec-gluon},
except that the q and $\overline{\mathrm{q}}$
for the {\gincl} jet selection are identified using parton
level information as described in Sect.~\ref{sec-gluon}
rather than using displaced secondary vertices:
we do not employ this latter method to obtain
the model predictions for
Figs.~\ref{fig-uncorry}b and~\ref{fig-uncorrnch}b
due to a lack of sufficient Ariadne Monte Carlo
event statistics which include simulation of the detector.
The data in 
Figs.~\ref{fig-uncorry}b and~\ref{fig-uncorrnch}b
have been corrected for the 18\% background to the
{\gincl} jets using bin-by-bin factors given by the
ratios of the Jetset predictions in
Figs.~\ref{fig-uncorry}b and~\ref{fig-uncorrnch}b
to those in
Figs.~\ref{fig-uncorry}a and~\ref{fig-uncorrnch}a,
respectively.
Thus the data and model results in
Figs.~\ref{fig-uncorry}b and~\ref{fig-uncorrnch}b
correspond to pure gluon jets which have not been
corrected for detector acceptance and resolution.
By comparing the relative differences between Jetset and
the data in Fig.~\ref{fig-uncorry}a and~b,
and similarly in
Fig.~\ref{fig-uncorrnch}a and~b,
it is seen that no significant bias is introduced in the
gluon jet measurements by applying the corrections
for background.

The discrepancies of {\artwo} and {\arthree} with the data,
noted above in connection with the fully corrected results
(cf.~Figs.~\ref{fig-rapidity} and~\ref{fig-nchy}),
are clearly visible in
Figs.~\ref{fig-uncorry}b and~\ref{fig-uncorrnch}b:
these two models predict significantly fewer particles
at small rapidities (\mbox{$\mmysph\leq 2$})
than are observed experimentally.
In contrast,
Jetset and the version of Ariadne without reconnection
are seen to describe the data well.
The $\chi^2$ values between the data and models
are 26, 17, 45 and 63 for Jetset, Ariadne, {\artwo}
and {\arthree},
for the 25 bins of data shown
in Fig.~\ref{fig-uncorry}b.
The corresponding results 
for the 10 bins of data with \mbox{$\mmysph\leq 2$}
are 6, 3, 29 and~43.
Integrating the distributions of
Fig.~\ref{fig-uncorrnch}b between
{\nch}$\,$=$\,$0 and {\nch}$\,$=$\,$5
(i.e.~similar to the test presented in Table~\ref{tab-nchle5}
for the fully corrected data),
we obtain $12.9\pm0.6\,$(stat.) for Jetset, 
$12.5\pm0.6\,$(stat.) for Ariadne, 
$20.1\pm0.6\,$(stat.) for {\artwo} and
$21.6\pm0.6\,$(stat.) for {\arthree},
compared to the measured value of
$11.4\pm1.8\,$(stat.):
this represents a discrepancy between data and model
of 4.7 standard deviations for {\artwo} and
of 5.4 standard deviations for~{\arthree}.
The value {\nch}$\,$=$\,$5 is chosen 
as the upper limit of integration for this last result
because it yields the maximum deviation of the predictions of
{\artwo} and {\arthree} with respect to
the standard version of Ariadne,
at the level including detector simulation,
compared to other choices.

On the basis of the results presented above,
we conclude that the {\artwo} and {\arthree} color reconnection
models implemented in Ariadne are disfavored.
This result may be of some benefit in the assessment
of systematic uncertainties for the
W boson mass measurement at LEP-2.


\section{Summary and conclusions}
\label{sec-summary}

In this paper,
we have presented experimental measurements
of the properties
of gluon and light flavored (uds) quark jets.
The jets are defined by inclusive
sums over the particles in {\gincl} and uds event
hemispheres,
with the {\gincl} gluon jet opposite to a hemisphere 
containing two identified quark jets in {\epem} annihilations
(the quark jets for the {\gincl} identification
are defined using a jet finding algorithm).
These inclusive definitions are in close
correspondence to the definition of jets used for
QCD calculations,
based on the production of virtual gluon and
quark jet pairs,
{\gluglu} and {\qq},
from a color singlet point source.
We present the distributions of rapidity,
scaled energy,
the logarithm of the momentum,
transverse momentum with respect to the jet axis,
and multiplicity in restricted intervals of rapidity,
for charged particles in the gluon and quark jets.
Our results for gluon jets
are almost entirely independent
of the choice of a jet finding algorithm,
a unique feature of our analysis
compared to other studies of high energy
({\ejet}$\,>\,$~5~GeV) gluon jets.
The energy of the jets in our study is about~40~GeV.

We determine the ratio, {$\rch$},
of the mean gluon to quark
jet charged particle multiplicity for
particles with rapidities 
{\mysph}$\,\leq\,$1 to be
$\rchysphone$$\,$=$\,$$1.919\pm 0.047\,(\mathrm{stat.})
\pm 0.095\,(\mathrm{syst.})$.
The corresponding ratio for soft particles at large
transverse momentum,
defined by $p$$\,<\,$4~GeV/$c$ and 
0.8$\,<\,$$\pperp$$\,<\,$3.0~GeV/$c$,
is found to be
$\rchsoft$$\,$=$\,$$2.29\pm 0.09\,(\mathrm{stat.})
\pm 0.15\,(\mathrm{syst.})$.
Our measurement of this last quantity is motivated
by the prediction that the multiplicity difference
between gluon and quark jets for soft particles emitted
at large angles to the jet axes approximately
equals the ratio of QCD color factors, {\cacf}$\,$=$\,$2.25,
even at the finite energies of LEP~\cite{bib-ochs}.
Using the Herwig Monte Carlo,
we verify that our results are
consistent with the QCD prediction
that the mean multiplicities of soft particles
in gluon and quark jets
differ by a factor of $r$$\,$=$\,${\cacf}$\,$=$\,$2.25~\cite{bib-stan},
once the effects of hadronization
and finite energy
have been considered.
Because our experimental definition of jets corresponds to
the theoretical one,
our results are the most direct test of this
QCD prediction to date.

Further,
we use our data to perform the most stringent test to date
of the model of color reconnection~\cite{bib-lonnblad}
implemented in the Ariadne Monte Carlo.
We find that this model
does not describe our gluon jet measurements accurately.
This result may be of some utility in assessing the
systematic uncertainty associated with color reconnection
in the determination
of the W boson mass from
{\epem}$\rightarrow$
W$^+$W$^-\rightarrow\,$
q$_1$$\overline{\mathrm{q}}_2$q$_3$$\overline{\mathrm{q}}_4$
events recorded at LEP-2.

\section{Acknowledgements}

We thank Stan Brodsky, Valery Khoze and Wolfgang Ochs
for valuable discussions,
and Torbj\"{o}rn Sj\"{o}strand for help in implementing
the version of Jetset with {\cca}$\,$=$\,${\ccf}
mentioned in Sect.~\ref{sec-mc}.

\noindent
We particularly wish to thank the SL Division for the 
efficient operation of the LEP accelerator
and for their continuing close cooperation with
our experimental group.  
We thank our colleagues from CEA, DAPNIA/SPP,
CE-Saclay for their efforts over the years on the time-of-flight and trigger
systems which we continue to use.  
In addition to the support staff at our own
institutions we are pleased to acknowledge the  \\
Department of Energy, USA, \\
National Science Foundation, USA, \\
Particle Physics and Astronomy Research Council, UK, \\
Natural Sciences and Engineering Research Council, Canada, \\
Israel Science Foundation, administered by the Israel
Academy of Science and Humanities, \\
Minerva Gesellschaft, \\
Benoziyo Center for High Energy Physics,\\
Japanese Ministry of Education, Science and Culture (the
Monbusho) and a grant under the Monbusho International
Science Research Program,\\
Japanese Society for the Promotion of Science (JSPS),\\
German Israeli Bi-national Science Foundation (GIF), \\
Bundesministerium f\"ur Bildung, Wissenschaft,
Forschung und Technologie, Germany, \\
National Research Council of Canada, \\
Research Corporation, USA,\\
Hungarian Foundation for Scientific Research, OTKA T-016660, 
T023793 and OTKA F-023259.\\

\pagebreak\clearpage

\pagebreak
\begin{table}[p]
\centering
\begin{tabular}{|c|ccc|}
 \hline
  & & & \\[-2.4mm]
$|y|$ & {\gincl} gluon jet & uds quark jet & Ratio \\[2mm]
 \hline
 \hline
0.0-0.2 & $7.42\pm0.35\pm0.70$ & $3.937\pm0.021\pm0.074$ & $1.97\pm0.09\pm0.20$ \\
0.2-0.4 & $6.37\pm0.31\pm0.59$ & $3.361\pm0.016\pm0.077$ & $1.97\pm0.10\pm0.18$ \\
0.4-0.6 & $5.86\pm0.26\pm0.48$ & $3.206\pm0.014\pm0.068$ & $1.90\pm0.09\pm0.17$ \\
0.6-0.8 & $5.39\pm0.29\pm0.61$ & $3.117\pm0.015\pm0.064$ & $1.80\pm0.10\pm0.21$ \\
0.8-1.0 & $5.31\pm0.26\pm0.61$ & $3.040\pm0.012\pm0.051$ & $1.82\pm0.09\pm0.21$ \\
1.0-1.2 & $5.40\pm0.28\pm0.65$ & $2.928\pm0.013\pm0.043$ & $1.92\pm0.10\pm0.23$ \\
1.2-1.4 & $5.30\pm0.25\pm0.54$ & $2.858\pm0.012\pm0.034$ & $1.93\pm0.09\pm0.20$ \\
1.4-1.6 & $4.86\pm0.21\pm0.52$ & $2.739\pm0.014\pm0.034$ & $1.84\pm0.08\pm0.20$ \\
1.6-1.8 & $4.39\pm0.20\pm0.42$ & $2.649\pm0.012\pm0.030$ & $1.72\pm0.08\pm0.17$ \\
1.8-2.0 & $3.98\pm0.21\pm0.32$ & $2.571\pm0.012\pm0.026$ & $1.60\pm0.09\pm0.13$ \\
2.0-2.2 & $3.47\pm0.21\pm0.28$ & $2.464\pm0.013\pm0.023$ & $1.46\pm0.09\pm0.12$ \\
2.2-2.4 & $3.06\pm0.21\pm0.23$ & $2.306\pm0.010\pm0.027$ & $1.38\pm0.10\pm0.11$ \\
2.4-2.6 & $2.85\pm0.18\pm0.22$ & $2.159\pm0.012\pm0.031$ & $1.38\pm0.09\pm0.11$ \\
2.6-2.8 & $2.00\pm0.18\pm0.28$ & $2.019\pm0.010\pm0.022$ & $1.04\pm0.09\pm0.15$ \\
2.8-3.0 & $1.84\pm0.13\pm0.24$ & $1.856\pm0.010\pm0.016$ & $1.05\pm0.08\pm0.14$ \\
3.0-3.2 & $1.22\pm0.13\pm0.25$ & $1.656\pm0.009\pm0.012$ & $0.79\pm0.08\pm0.16$ \\
3.2-3.4 & $0.96\pm0.11\pm0.16$ & $1.457\pm0.008\pm0.013$ & $0.72\pm0.08\pm0.12$ \\
3.4-3.6 & $0.63\pm0.09\pm0.17$ & $1.243\pm0.008\pm0.020$ & $0.56\pm0.08\pm0.16$ \\
3.6-3.8 & $0.62\pm0.08\pm0.12$ & $1.015\pm0.007\pm0.021$ & $0.69\pm0.08\pm0.13$ \\
3.8-4.0 & $0.32\pm0.06\pm0.11$ & $0.834\pm0.007\pm0.024$ & $0.43\pm0.09\pm0.15$ \\
4.0-4.2 & $0.185\pm0.043\pm0.076$ & $0.658\pm0.005\pm0.017$ & $0.32\pm0.08\pm0.13$ \\
4.2-4.4 & $0.169\pm0.047\pm0.070$ & $0.486\pm0.005\pm0.018$ & $0.41\pm 0.11\pm0.16$ \\
4.4-4.6 & $0.100\pm0.034\pm0.046$ & $0.361\pm0.004\pm0.020$ & $0.32\pm0.11\pm0.15$ \\
4.6-4.8 & $0.046\pm0.026\pm0.046$ & $0.265\pm0.004\pm0.020$ & $0.21\pm0.11\pm0.20$ \\
4.8-5.0 & $0.048\pm0.020\pm0.031$ & $0.197\pm0.003\pm0.019$ & $0.29\pm0.13\pm0.19$ \\
5.0-5.2 & $0.013\pm0.018\pm0.023$ & $0.133\pm0.002\pm0.015$ & $0.12\pm0.17\pm0.18$ \\
5.2-5.4 & ---  & $0.09549\pm0.0026\pm0.0098$ & ---  \\
5.4-5.6 & $0.0051\pm0.0082\pm0.0055$ & $0.0626\pm0.0019\pm0.0054$ & $0.10\pm0.15\pm0.09$ \\
5.6-5.8 & ---& $0.0453\pm0.0014\pm0.0034$ & --- \\
5.8-6.0 & ---& $0.0290\pm0.0012\pm0.0020$ & --- \\
 \hline
\end{tabular}
\caption{The charged particle rapidity distribution, $|y|$,
of 40.1~GeV {\gincl}
gluon jets and 45.6~GeV uds quark jets, 
and the ratio of 40.1~GeV {\gincl} to 40.1~GeV uds quark jets.
The first uncertainty is statistical and the second is systematic.
These data are displayed in Fig.~\ref{fig-rapidity}.
}
\label{tab-rapidity}
\end{table}

\pagebreak
\begin{table}[p]
\centering
\begin{tabular}{|c|ccc|}
 \hline
  & & & \\[-2.4mm]
{\xee} & {\gincl} gluon jet & uds quark jet & Ratio \\[2mm]
 \hline
 \hline
0.00-0.01 & $248\pm9\pm21$       & $170.1\pm0.5\pm1.7$     & $1.80\pm0.06\pm0.13$ \\
0.01-0.02 & $328\pm10\pm18$      & $196.7\pm0.6\pm1.5$     & $1.79\pm0.05\pm0.12$ \\
0.02-0.03 & $222\pm8\pm16$       & $128.7\pm0.4\pm1.4$     & $1.80\pm0.06\pm0.12$ \\
0.03-0.04 & $155\pm6\pm11$       & $87.7\pm0.3\pm1.1$      & $1.79\pm0.07\pm0.13$ \\
0.04-0.05 & $97.7\pm5.0\pm8.6$   & $63.84\pm0.25\pm0.92$   & $1.54\pm0.08\pm0.13$ \\
0.05-0.06 & $67.9\pm4.0\pm7.9$   & $48.31\pm0.24\pm0.58$   & $1.41\pm0.08\pm0.17$ \\
0.06-0.07 & $55.4\pm3.6\pm6.2$   & $38.44\pm0.16\pm0.50$   & $1.45\pm0.09\pm0.16$ \\
0.07-0.08 & $40.1\pm2.8\pm6.0$   & $31.22\pm0.18\pm0.42$   & $1.28\pm0.09\pm0.18$ \\
0.08-0.09 & $38.2\pm2.6\pm4.7$   & $26.37\pm0.17\pm0.47$   & $1.44\pm0.10\pm0.18$ \\
0.09-0.10 & $28.5\pm2.4\pm4.0$   & $22.04\pm0.13\pm0.28$   & $1.29\pm0.11\pm0.18$ \\
0.10-0.12 & $23.0\pm1.7\pm3.6$   & $17.79\pm0.09\pm0.22$   & $1.29\pm0.09\pm0.20$ \\
0.12-0.14 & $14.9\pm1.2\pm2.3$   & $13.238\pm0.075\pm0.096$ & $1.11\pm0.09\pm0.17$ \\
0.14-0.16 & $10.0\pm1.0\pm1.8$   & $10.44\pm0.08\pm0.11$   & $0.95\pm0.10\pm0.16$ \\
0.16-0.18 & $7.7\pm0.9\pm1.3$    & $8.231\pm0.061\pm0.092$ & $0.94\pm0.11\pm0.15$ \\
0.18-0.20 & $4.94\pm0.76\pm0.86$ & $6.776\pm0.056\pm0.086$ & $0.72\pm0.11\pm0.13$ \\
0.20-0.25 & $3.15\pm0.37\pm0.77$ & $4.829\pm0.035\pm0.053$ & $0.64\pm0.08\pm0.14$ \\
0.25-0.30 & $1.85\pm0.27\pm0.38$ & $3.105\pm0.022\pm0.049$ & $0.58\pm0.09\pm0.12$ \\
0.30-0.40 & $0.52\pm0.11\pm0.24$ & $1.655\pm0.012\pm0.039$ & $0.31\pm0.07\pm0.11$ \\
0.40-0.50 & $0.14\pm0.06\pm0.10$ & $0.757\pm0.007\pm0.036$ & $0.184\pm0.075\pm0.084$ \\
0.50-0.60 & $0.019\pm0.015\pm0.032$ & $0.339\pm0.006\pm0.017$ & $0.056\pm0.043\pm0.087$ \\
0.60-0.80 & $0.014\pm0.011\pm0.012$ & $0.1118\pm0.0026\pm0.0070$ & $0.121\pm0.093\pm0.078$ \\
0.80-1.00 & $0.0005\pm0.0004\pm0.008$ & $0.0143\pm0.0009\pm0.0026$ & $0.030\pm0.026\pm0.092$ \\
 \hline
\end{tabular}
\caption{The charged particle scaled energy, {\xee}$\,$=$\,$$E/E_{\,\mathrm{jet}}$,
of 40.1~GeV {\gincl}
gluon jets and 45.6~GeV uds quark jets, 
and the ratio of 40.1~GeV {\gincl} to 40.1~GeV uds quark jets.
The first uncertainty is statistical and the second is systematic.
These data are displayed in Fig.~\ref{fig-xee}.
}
\label{tab-xee}
\end{table}

\pagebreak
\begin{table}[p]
\centering
\begin{tabular}{|c|ccc|}
 \hline
  & & & \\[-2.4mm]
\mbox{$\ln\,(p)$} & {\gincl} gluon jet & uds quark jet & Ratio \\[2mm]
 \hline
 \hline
--2.5 - --2.0 & $0.48\pm0.05\pm0.60$ & $0.27\pm0.01\pm0.35$ & $1.8\pm0.2\pm1.2$ \\
--2.0 - --1.5 & $1.33\pm0.08\pm0.67$ & $0.78\pm0.01\pm0.35$ & $1.7\pm0.1\pm1.2$ \\
--1.5 - --1.0 & $2.84\pm0.12\pm0.64$ & $1.58\pm0.01\pm0.26$ & $1.83\pm0.08\pm0.67$ \\
--1.0 - --0.5 & $4.20\pm0.15\pm0.36$ & $2.431\pm0.008\pm0.021$ & $1.76\pm0.06\pm0.17$ \\
--0.5 - 0.0 & $5.17\pm0.15\pm0.33$     & $2.925\pm0.009\pm0.026$ & $1.81\pm0.05\pm0.13$ \\
0.0 - 0.5   & $5.07\pm0.15\pm0.22$       & $3.012\pm0.008\pm0.028$ & $1.749\pm0.053\pm0.084$ \\
0.5 - 1.0   & $3.90\pm0.12\pm0.20$       & $2.749\pm0.009\pm0.029$ & $1.495\pm0.047\pm0.080$ \\
1.0 - 1.5   & $3.11\pm0.10\pm0.18$       & $2.299\pm0.007\pm0.023$ & $1.447\pm0.048\pm0.087$ \\
1.5 - 2.0   & $1.67\pm0.08\pm0.13$       & $1.757\pm0.006\pm0.016$ & $1.043\pm0.052\pm0.073$ \\
2.0 - 2.5   & $0.670\pm0.050\pm0.078$  & $1.167\pm0.005\pm0.012$ & $0.653\pm0.049\pm0.083$ \\
2.5 - 3.0   & $0.091\pm0.024\pm0.036$  & $0.592\pm0.003\pm0.012$ & $0.192\pm0.051\pm0.080$ \\
3.0 - 3.5   & $0.004\pm0.003\pm0.023$  & $0.1788\pm0.0017\pm0.0083$ & $0.032\pm0.023\pm0.093$ \\
3.5 - 4.0   & $0.0002\pm0.0002\pm0.0046$ & $0.0138\pm0.0006\pm0.0028$ & $0.041\pm0.032\pm0.059$ \\
 \hline
\end{tabular}
\caption{The logarithm of charged particle momentum, \mbox{$\ln\,(p)$},
of 40.1~GeV {\gincl}
gluon jets and 45.6~GeV uds quark jets, 
and the ratio of 40.1~GeV {\gincl} to 40.1~GeV uds quark jets.
The first uncertainty is statistical and the second is systematic.
These data are displayed in Fig.~\ref{fig-lnp}.
}
\label{tab-lnp}
\end{table}

\pagebreak
\begin{table}[p]
\centering
\begin{tabular}{|c|ccc|}
 \hline
  & & & \\[-2.4mm]
$\pperp$ & {\gincl} gluon jet & uds quark jet & Ratio \\[2mm]
 \hline
 \hline
0.0-0.1 & $9.73\pm0.47\pm0.92$  & $7.38\pm0.03\pm0.24$    
  & $1.359\pm0.065\pm0.093$ \\
0.1-0.2 & $21.3\pm0.6\pm1.0$    & $15.16\pm0.04\pm0.19$   
  & $1.445\pm0.044\pm0.085$ \\
0.2-0.3 & $20.9\pm0.8\pm1.2$    & $15.72\pm0.04\pm0.15$   
  & $1.369\pm0.051\pm0.081$ \\
0.3-0.4 & $17.3\pm0.6\pm1.1$    & $13.02\pm0.04\pm0.10$   
  & $1.377\pm0.050\pm0.093$ \\
0.4-0.5 & $14.04\pm0.58\pm0.97$ & $10.02\pm0.03\pm0.11$   
  & $1.46\pm0.06\pm0.10$ \\
0.5-0.6 & $11.37\pm0.50\pm0.85$ & $7.571\pm0.024\pm0.098$ 
  & $1.57\pm0.07\pm0.13$ \\
0.6-0.7 & $8.82\pm0.45\pm0.92$  & $5.758\pm0.022\pm0.068$ 
  & $1.62\pm0.08\pm0.18$ \\
0.7-0.8 & $7.67\pm0.40\pm0.90$  & $4.391\pm0.024\pm0.052$ 
  & $1.87\pm0.10\pm0.21$ \\
0.8-1.0 & $5.42\pm0.23\pm0.56$  & $2.989\pm0.012\pm0.042$ 
  & $1.97\pm0.08\pm0.18$ \\
1.0-1.4 & $2.64\pm0.13\pm0.37$  & $1.590\pm0.007\pm0.032$ 
  & $1.84\pm0.09\pm0.30$ \\
1.4-2.0 & $1.24\pm0.07\pm0.19$  & $0.680\pm0.004\pm0.013$ 
  & $2.08\pm0.12\pm0.34$ \\
2.0-3.0 & $0.36\pm0.03\pm0.15$  & $0.2497\pm0.0018\pm0.0054$ 
  & $1.71\pm0.16\pm0.53$ \\
3.0-4.0 & $0.109\pm0.017\pm0.041$ & $0.0903\pm0.0010\pm0.0028$ 
  & $1.51\pm0.23\pm0.46$ \\
4.0-6.0 & $0.029\pm0.005\pm0.022$ & $0.0289\pm0.0004\pm0.0015$ 
  & $1.34\pm0.22\pm0.73$ \\
6.0-8.0 & $0.0036\pm0.0015\pm0.0065$ & $0.00860\pm0.00016\pm0.00083$ 
  & $0.60\pm0.25\pm1.14$ \\
8.0-10.0 & $0.0035\pm0.0002\pm0.0041$ & $0.00304\pm0.00010\pm0.00051$ 
  & $1.75\pm0.11\pm1.50$ \\
  \hline
\end{tabular}
\caption{The charged particle transverse momentum with
respect to the jet axis, $\pperp$,
of 40.1~GeV {\gincl}
gluon jets and 45.6~GeV uds quark jets, 
and the ratio of 40.1~GeV {\gincl} to 40.1~GeV uds quark jets.
The first uncertainty is statistical and the second is systematic.
These data are displayed in Fig.~\ref{fig-ptall}.
}
\label{tab-ptall}
\end{table}

\pagebreak
\begin{table}[p]
\centering
\begin{tabular}{|c|ccc|}
 \hline
  & & & \\[-2.4mm]
$\ptsoft$ & {\gincl} gluon jet & uds quark jet & Ratio \\[2mm]
 \hline
 \hline
0.0-0.1 & $9.50\pm0.46\pm0.97$ & $6.91\pm0.03\pm0.22$  
  & $1.41\pm0.07\pm0.13$ \\
0.1-0.2 & $20.8\pm0.6\pm1.0$ & $13.85\pm0.04\pm0.16$   
  & $1.53\pm0.05\pm0.11$ \\
0.2-0.3 & $20.2\pm0.8\pm1.2$ & $13.86\pm0.04\pm0.13$   
  & $1.49\pm0.06\pm0.11$ \\
0.3-0.4 & $16.6\pm0.6\pm1.2$ & $10.94\pm0.04\pm0.10$   
  & $1.55\pm0.06\pm0.13$ \\
0.4-0.5 & $13.3\pm0.6\pm1.1$ & $8.049\pm0.03\pm0.11$   
  & $1.70\pm0.07\pm0.16$ \\
0.5-0.6 & $10.1\pm0.5\pm1.1$ & $5.836\pm0.023\pm0.091$ 
  & $1.79\pm0.09\pm0.23$ \\
0.6-0.7 & $7.7\pm0.4\pm1.0$  & $4.301\pm0.020\pm0.068$ 
  & $1.86\pm0.11\pm0.26$ \\
0.7-0.8 & $6.66\pm0.38\pm0.90$ & $3.169\pm0.021\pm0.052$ 
  & $2.21\pm0.13\pm0.30$ \\
0.8-1.0 & $4.51\pm0.22\pm0.48$ & $2.051\pm0.011\pm0.044$ 
  & $2.35\pm0.12\pm0.25$ \\
1.0-1.4 & $2.07\pm0.13\pm0.33$ & $1.031\pm0.006\pm0.032$ 
  & $2.18\pm0.13\pm0.42$ \\
1.4-2.0 & $0.91\pm0.06\pm0.15$ & $0.3876\pm0.0030\pm0.0095$ 
  & $2.62\pm0.17\pm0.46$ \\
2.0-3.0 & $0.16\pm0.03\pm0.10$ & $0.1074\pm0.0012\pm0.0037$ 
  & $1.71\pm0.28\pm0.82$ \\
3.0-4.0 & $0.0167\pm0.0084\pm0.019$ & $0.01837\pm0.00047\pm0.00098$ 
  & $1.11\pm0.56\pm0.81$ \\
 \hline
\end{tabular}
\caption{The charged particle transverse momentum with
respect to the jet axis
for particles with momentum $p$$\,<\,$4~GeV/$c$, $\ptsoft$,
of 40.1~GeV {\gincl}
gluon jets and 45.6~GeV uds quark jets, 
and the ratio of 40.1~GeV {\gincl} to 40.1~GeV uds quark jets.
The first uncertainty is statistical and the second is systematic.
These data are displayed in Fig.~\ref{fig-ptsoft}.
}
\label{tab-ptsoft}
\end{table}

\pagebreak
\begin{table}[p]
\centering
\begin{tabular}{|c|cc|}
 \hline
  & & \\[-2.4mm]
$n_{\,\mathrm{ch.}}(\mmysph\leq 2)$
 & {\gincl} gluon jet & uds quark jet \\[2mm]
 \hline
 \hline
 0 &  $0.15\pm0.01\pm0.26$ & $2.13\pm0.04\pm0.13$ \\
 1 &  $0.34\pm0.23\pm0.31$ & $5.83\pm0.07\pm0.17$ \\
 2 &  $0.94\pm0.47\pm0.40$ & $9.93\pm0.08\pm0.22$ \\
 3 &  $2.13\pm0.74\pm0.56$ & $12.37\pm0.10\pm0.21$ \\
 4 &  $2.86\pm0.97\pm0.83$ & $12.67\pm0.10\pm0.21$ \\
 5 &  $4.0\pm1.1\pm1.2$    &  $11.47\pm0.09\pm0.19$ \\
 6 &  $6.0\pm1.2\pm1.4$    &  $9.52\pm0.10\pm0.18$ \\
 7 &  $7.6\pm1.1\pm1.2$    &  $7.63\pm0.07\pm0.18$ \\
 8 &  $8.9\pm1.3\pm0.83$  &  $5.98\pm0.08\pm0.17$ \\
 9 &  $9.7\pm1.2\pm0.8$    &  $4.74\pm0.07\pm0.16$ \\
10 &  $9.9\pm1.1\pm0.8$    &  $3.75\pm0.05\pm0.13$ \\
11 &  $9.0\pm1.4\pm1.0$    &  $3.02\pm0.06\pm0.10$ \\
12 &  $7.9\pm1.2\pm1.0$    &  $2.418\pm0.042\pm0.072$ \\
13 &  $6.5\pm1.2\pm0.9$    &  $1.937\pm0.045\pm0.052$ \\
14 &  $5.2\pm1.2\pm0.8$    &  $1.561\pm0.034\pm0.043$ \\
15 &  $4.04\pm0.85\pm0.72$ & $1.218\pm0.036\pm0.038$ \\
16 &  $3.05\pm0.90\pm0.55$ & $0.957\pm0.029\pm0.038$ \\
17 &  $2.64\pm0.77\pm0.44$ & $0.734\pm0.027\pm0.036$ \\
18 &  $1.84\pm0.66\pm0.74$ & $0.564\pm0.022\pm0.033$ \\
19 &  $1.60\pm0.68\pm0.78$ & $0.423\pm0.019\pm0.027$ \\
20 &  $1.45\pm0.53\pm0.86$ & $0.322\pm0.017\pm0.020$ \\
21 &  $0.75\pm0.43\pm0.55$ & $0.241\pm0.014\pm0.014$ \\
22 &  $0.93\pm0.40\pm0.62$ & $0.181\pm0.011\pm0.011$ \\
23 &  $0.95\pm0.39\pm0.42$ & $0.132\pm0.010\pm0.010$ \\
24 &  $0.52\pm0.27\pm0.56$ & $0.0894\pm0.0090\pm0.0073$ \\
25 &  $0.24\pm0.21\pm0.48$ & $0.0640\pm0.0081\pm0.0053$ \\
26 &  $0.19\pm0.15\pm0.38$ & $0.0440\pm0.0059\pm0.0047$ \\
27 &  ---    &  $0.0267\pm0.0047\pm0.0052$ \\
28 &  $0.25\pm0.09\pm0.27$ & $0.0207\pm0.0039\pm0.0056$ \\
29 &  $0.19\pm0.06\pm0.30$ & $0.0144\pm0.0027\pm0.0040$ \\
30 &  $0.09\pm0.05\pm0.13$ & $0.0084\pm0.0023\pm0.0025$ \\
31 &  ---    &  $0.0036\pm0.0020\pm0.0024$ \\
32 &  ---    &  $0.0031\pm0.0014\pm0.0016$ \\
33 &  ---    &  $0.0007\pm0.0011\pm0.0015$ \\
34 &  ---    &  --- \\
35 &  ---    &  $0.00079\pm0.00056\pm0.00065$ \\
36 &  ---    &  $0.00031\pm0.00034\pm0.00045$ \\
 \hline
  & & \\[-2.4mm]
$\langle n_{\,\mathrm{ch.}}(\mmysph\leq 2)\rangle$
 &  $10.83\pm0.20\pm0.41$    &  $6.085\pm0.013\pm0.071$ \\[2mm]
  \hline
\end{tabular}
\caption{Charged particle multiplicity distributions,
expressed in per cent (\%),
of 40.1~GeV {\gincl} gluon jets and 45.6~GeV uds quark jets,
in the rapidity interval \mbox{{\mysph}$\,\leq\,$2}.
The mean values 
$\langle n_{\,\mathrm{ch.}}(\mmysph\leq 2)\rangle$
are also given.
The first uncertainty is statistical and the second is systematic.
The data are correlated between bins.
These data are displayed in Fig.~\ref{fig-nchy}a.
}
\label{tab-nchylttwo}
\end{table}

\pagebreak
\begin{table}[p]
\centering
\begin{tabular}{|c|cc|}
 \hline
  & & \\[-2.4mm]
$n_{\,\mathrm{ch.}}(\mmysph\leq 1)$
 & {\gincl} gluon jet & uds quark jet \\[2mm]
 \hline
 \hline
0 &  $1.3\pm0.5\pm1.1$       &  $11.25\pm0.09\pm0.41$ \\
1 &  $4.5\pm0.9\pm1.2$       &  $18.77\pm0.11\pm0.33$ \\
2 &  $7.7\pm1.3\pm1.2$       &  $19.16\pm0.12\pm0.32$ \\
3 &  $10.9\pm1.5\pm1.1$     &  $15.23\pm0.10\pm0.26$ \\
4 &  $11.9\pm1.6\pm0.9$     &  $10.66\pm0.09\pm0.26$ \\
5 &  $12.8\pm1.6\pm1.0$     &  $7.16\pm0.07\pm0.22$ \\
6 &  $11.9\pm1.6\pm1.1$     &  $4.92\pm0.06\pm0.20$ \\
7 &  $10.3\pm1.5\pm1.4$     &  $3.44\pm0.06\pm0.16$ \\
8 &  $7.8\pm1.2\pm1.3$       &  $2.56\pm0.04\pm0.12$ \\
9 &  $5.7\pm1.2\pm1.0$       &  $1.897\pm0.042\pm0.079$ \\
10 &  $3.79\pm0.98\pm0.71$ &  $1.395\pm0.040\pm0.064$ \\
11 &  $2.92\pm0.92\pm0.64$ &  $1.032\pm0.032\pm0.052$ \\
12 &  $2.16\pm0.74\pm0.60$ &  $0.750\pm0.024\pm0.039$ \\
13 &  $1.57\pm0.65\pm0.91$ &  $0.546\pm0.023\pm0.025$ \\
14 &  $1.52\pm0.54\pm0.89$ &  $0.395\pm0.021\pm0.014$ \\
15 &  $0.72\pm0.48\pm0.77$ &  $0.273\pm0.017\pm0.010$ \\
16 &  $0.71\pm0.37\pm0.53$ &  $0.194\pm0.013\pm0.008$ \\
17 &  $0.64\pm0.25\pm0.59$ &  $0.128\pm0.011\pm0.006$ \\
18 &  $0.56\pm0.21\pm0.64$ &  $0.0847\pm0.0087\pm0.0049$ \\
19 &  $0.36\pm0.21\pm0.54$ &  $0.0555\pm0.0072\pm0.0041$ \\
20 &  $0.25\pm0.12\pm0.28$ &  $0.0376\pm0.0047\pm0.0036$ \\
21 &  ---    &  $0.0204\pm0.0045\pm0.0038$ \\
22 &  ---    &  $0.0136\pm0.0036\pm0.0033$ \\
23 &  ---    &  $0.0101\pm0.0026\pm0.0041$ \\
24 &  ---    &  $0.0058\pm0.0019\pm0.0041$ \\
25 &  ---    &  $0.0036\pm0.0015\pm0.0041$ \\
26 &  ---    &  $0.0022\pm0.0011\pm0.0023$ \\
 \hline
  & & \\[-2.4mm]
$\langle n_{\,\mathrm{ch.}}(\mmysph\leq 1)\rangle$
 &  $6.14\pm0.15\pm0.36$    &  $3.333\pm0.010\pm0.046$ \\[2mm]
  \hline
\end{tabular}
\caption{Charged particle multiplicity distributions,
expressed in per cent (\%),
of 40.1~GeV {\gincl} gluon jets and 45.6~GeV uds quark jets,
in the rapidity interval \mbox{{\mysph}$\,\leq\,$1}.
The mean values 
$\langle n_{\,\mathrm{ch.}}(\mmysph\leq 1)\rangle$
are also given.
The first uncertainty is statistical and the second is systematic.
The data are correlated between bins.
These data are displayed in Fig.~\ref{fig-nchy}b.
}
\label{tab-nchyltone}
\end{table}

\begin{table}[htbp]
\centering
\begin{tabular}{|c|cccc|}
 \hline
  & & & & \\[-2.4mm]
  & {$\rch$} & {$\rchysphtwo$} & {$\rchysphone$} & $\rchsoft$ \\[2mm]
 \hline
 \hline
OPAL data & $1.514\pm0.019$  & $1.852\pm0.034$
      & $1.919\pm0.047$ &  $2.29\pm0.09$ \\
   & $\pm0.034$   &  $\pm0.077$  &  $\pm0.095$ &  $\pm 0.15$ \\
 \hline
 \hline
Standard QCD models: & & & & \\
Herwig & 1.54 (1.56) & 1.85 (1.97) & 1.92 (2.06) &  2.16 (2.09) \\
Jetset  & 1.54 (1.35) & 1.84 (1.61) & 1.88 (1.62) & 1.93 (1.50) \\
Ariadne & 1.55 (1.46) & 1.89 (1.77) & 1.97 (1.81) & 2.07 (1.81) \\
 \hline
 \hline
Jetset {\cca}$\,$=$\,${\ccf}$\,$=$\,$4/3 & 1.38 (1.03) & 1.54 (1.06) & 1.51 (1.04) & 1.31 (1.00) \\
 \hline
 \hline
Reconnected QCD models: & & & & \\
{\artwo} & 1.43 & 1.72 & 1.78 & 1.95\\
{\arthree} & 1.42 & 1.69 & 1.75 & 1.92 \\
 \hline
\end{tabular}
\caption{
The ratios {$\rch$} of the mean charged particle multiplicity
between 40.1~GeV {\gincl} gluon jets
and 40.1~GeV uds quark jets for full phase space,
for restricted rapidity intervals $\mmysph\leq 2$
and $\mmysph\leq 1$,
and for soft particles at large transverse momentum
$\pperp$ with respect to the jet axis,
defined by $p$$\,<\,$4~GeV/$c$
and 0.8$\,<\,$$\pperp$$\,<\,$3~GeV/$c$,
with $p$ the particle momentum.
The results are given for the data and 
for QCD Monte Carlo programs.
The corresponding results at the parton level are given
in parentheses for the standard QCD models and for a
special version of Jetset with {\cca}$\,$=$\,${\ccf}$\,$=$\,$4/3.
For the data,
the first uncertainty is statistical and the second is systematic.
}
\label{tab-rch}
\end{table}

\begin{table}[htbp]
\centering
\begin{tabular}{|c|c|}
 \hline
  & \\[-2.4mm]
  & Probability for {\nch}($\mmysph\leq 2$)$\leq$4 \\
  & in {\gincl} jets \\[2mm]
 \hline
 \hline
OPAL data & $6.42\pm1.30\pm1.65$\%  \\
 \hline
 \hline
Standard QCD models: & \\
Jetset  & 6.0\% \\
Herwig & 7.1\% \\
Ariadne & 5.4\% \\
 \hline
 \hline
Reconnected QCD models: & \\
{\artwo} & 11.2\% \\
{\arthree} & 12.2\% \\
 \hline
\end{tabular}
\caption{
Probability,
measured in per cent,
for a {\gincl} jet to have four or fewer charged
particles with rapidity $\mmysph\leq 2$.
The results are given for the data and 
for QCD Monte Carlo programs.
For the data,
the first uncertainty is statistical and the second is systematic.
}
\label{tab-nchle5}
\end{table}

\pagebreak\clearpage

\clearpage

\pagebreak\clearpage

\clearpage

\pagebreak\clearpage

\clearpage

\clearpage

\clearpage

\clearpage

\clearpage

\clearpage

\clearpage

\clearpage

\pagebreak\clearpage

\clearpage

\clearpage

\clearpage

\clearpage

\clearpage

\end{document}